\documentclass[fleqn,10pt]{wlscirep}

\usepackage{amsmath,amssymb,esint}				
\usepackage{lmodern}							
\graphicspath{{./Figures/}}

\usepackage{lineno}

\newcommand{\be}{\begin{equation}}
\newcommand{\ee}{\end{equation}}
\newcommand{\BE}{\begin{eqnarray}}
\newcommand{\EE}{\end{eqnarray}}
\newcommand{\fbeta}{\mathcal{B}}
\newcommand{\fchi}{\mathfrak{X}}
\newcommand{\nn}{\nonumber}
\newcommand{\bra}{\left\langle}
\newcommand{\ket}{\right\rangle}
\newcommand{\avg}[1]{\bra{#1}\ket}

\title{The effects of heterogeneity on stochastic cycles in epidemics}

\author[1,*]{Francisco Herrer\'{i}as-Azcu\'{e}}
\author[1,+]{Tobias Galla}
\affil[1]{Theoretical Physics, School of Physics and Astronomy,
			The University of Manchester, Manchester M13 9PL, United Kingdom}

\affil[*]{francisco.herreriasazcue@postgrad.manchester.ac.uk}
\affil[+]{tobias.galla@manchester.ac.uk}

		\begin{abstract}

Models of biological processes are often subject to different sources of noise. Developing an understanding of the combined effects of different types of uncertainty is an open challenge. In this paper, we study a variant of the susceptible-infective-recovered model of epidemic spread, which combines both agent-to-agent heterogeneity and intrinsic noise.  We focus on epidemic cycles, driven by the stochasticity of infection and recovery events, and study in detail how heterogeneity in susceptibilities and propensities to pass on the disease affects these quasi-cycles. While the system can only be described by a large hierarchical set of equations in the transient regime, we derive a reduced closed set of equations for population-level quantities in the stationary regime. We analytically obtain the spectra of quasi-cycles in the linear-noise approximation. We find that the characteristic frequency of these cycles is typically determined by population averages of susceptibilities and infectivities, but that their amplitude depends on higher-order moments of the heterogeneity. We also investigate the synchronisation properties and phase lag between different groups of susceptible and infected individuals.

		\end{abstract}

\begin{document}

\flushbottom
\maketitle
\thispagestyle{empty}

		\section*{Introduction}

It is now widely recognised that noise and uncertainty play an important role in modelling biological systems. Traditional approaches to modelling phenomena in biology\cite{Murray2002} are often based on deterministic ordinary or partial differential equations, and do not aim to describe stochasticity. In order to capture epistemic uncertainty, static or dynamic noise variables are introduced in more modern mathematical biology.
This randomness reflects the lack of detailed knowledge about phenomena at finer scales than described by the model at hand; any modelling approach necessarily operates at a set scale (e.g. cell, individual, or population), and does not capture in detail the processes at smaller scales. These are `emulated' through effective randomness. Different types of such noise are frequently found in models of biological phenomena, including intrinsic demographic noise, extrinsic stochasticity, parameter uncertainty or heterogeneity between different types of interacting entities \cite{Wilkinson2011, Goel1974}. Some of these random variables are static and do not evolve in time, others are described by dynamic time-dependent noise. 
Intrinsic noise, due to the stochastic dynamics of a system has lately been the focus of many studies (see for example \cite{Andersson2000, Elowitz2011, Paulsson2004}). Extrinsic or parametric noise, due to variations, heterogeneity or uncertainties in the parameters or the environment surrounding the process, has received similar attention (e.g. \cite{Moreno2002, Raj2008}). To be able to adequately describe biological systems, however, it may be necessary to account for both these uncertainties which contribute to the noisy dynamics.

In the modelling of epidemics this is of particular importance. The infection process, driven by serendipitous contacts, is inherently stochastic, and heterogeneity in susceptibility to a disease or infectiousness of different individuals are known to exist and play a role in viral spread. For example, variation in host susceptibility and viral reproduction have been observed in \cite{Heldt2015}, and behavioural, structural or contact differences between individuals are inevitable. 
However, the better part of the existing work focusing on heterogeneity of this type, does not explicitly seek to capture demographic noise. Instead one often assumes infinite populations and deterministic dynamics. This approach is often taken outside epidemics as well. Much existing work studies {\em individual} sources of uncertainty, heterogeneity and noise in isolation, but not their interacting together.  A notable exception is the modelling of gene regulatory networks, in which the interaction of intrinsic and extrinsic noise is actively studied, see e.g. \cite{Scott2006, Swain2002, Hilfinger2011a}.

The effects of intrinsic noise have been recognised in recent years. In models with demographic processes, for example, intrinsic stochasticity has been seen to lead to sustained quasi-cycles \cite{Alonso2007, Olsen1990, Black2009, Rozhnova2009} in parameter regimes in which a deterministic model would converge to a stable fixed point. These quasi-cycles have been identified not only in models of epidemic spread, but also in other instances of population dynamics, including in genetic circuits, evolutionary systems and in game theory \cite{McKane2005a, Bjornstad2001, Bladon2010, Samoilov2005}. Heterogeneity has been and is being considered in epidemics as well. Age structure is studied for example in \cite{Bolker1993, Schenzle1984}, seasonally changing infection rates in \cite{Earn2000, Stone2007}, variation in infectivity and/or susceptibility are addressed in \cite{Diekmann1990, Hethcote1987, Nold1980, Hickson2014, Novozhilov2012}, spatial structure has been approached in \cite{Keeling1999, Rohani1999, Hagenaars2004, Yu2009}, and epidemics on static and dynamic networks are studied in \cite{Colizza2006, Barthelemy2004, Keeling2005, Moreno2002, Hufnagel2004, Holme2015}. Heterogeneity has been found to generate outbreaks that propagate hierarchically \cite{Barthelemy2004, Barthelemy2005}, grow faster than in homogeneous populations \cite{Keeling2005}, and have a lower total number of infected individuals \cite{Boylan1991, Andersson1998}.

Much of this work, whether describing a well-mixed population, a compartmented or structured one, is based on variants of the celebrated susceptible-infective-recovered (SIR) model. They can be described either by deterministic differential equations, or as a stochastic process involving a population of discrete individuals. In the former approach the population is effectively assumed to be infinite, so that the timing of stochastic infection, recovery or birth-death events `averages' out, and smooth laws for the time evolution of the population are obtained. The latter approach explicitly captures the intrinsic randomness of infection, recovery and demographics. The population is taken to be finite, and its state discrete. The model evolves through discrete events (e.g. infections). In the simplest case this defines a Markovian random process, which often can be analysed further mathematically, at least to a good approximation. Starting from the master equation in a well-mixed population a set of stochastic differential equations can be derived in the limit of large, but finite populations \cite{Gardiner2003}. These can then be studied further within the `linear-noise approximation' (LNA) \cite{vanKampen1992}. The mathematics are tractable and the corresponding theory is now well established. While remarkably powerful, this approach so far has mostly been used for well-mixed populations. The linear-noise approximation has also been applied to networked systems with contact heterogeneity (see e.g.  \cite{Rozhnova2009,Rozhnova2009a}), but progress is then much harder and often relies on further moment-closure approximations.

The aim of our work is to introduce agent-to-agent heterogeneity into the SIR dynamics in a finite well-mixed population. This provides a middle ground between homogeneous well-mixed models and an explicitly networked population. At the same time, we maintain tractability and are able to characterise stochastic effects in finite populations via the linear-noise approximation. This allows us to systematically investigate the combination of parameter heterogeneity and demographic noise.
We divide the population of agents into $K$ different groups of susceptible individuals, where members of different groups have different susceptibilities. Similarly, in our model there are $M$ classes of infective individuals, with each class representing a different propensity to pass on the disease. This follows the lines of \cite{Novozhilov2012}, but we explicitly focus on the combination of heterogeneity and intrinsic noise. Intrinsic stochasticity had not been included in  \cite{Novozhilov2012}.

Our paper is organised as follows: In Sec.~\ref{sec:Model} we describe our model in detail. As a baseline we then construct the deterministic rate equations in Sec.~\ref{sec:Deterministic}. They describe the deterministic dynamics in the limit of infinite populations, and are required to carry out the LNA. The most natural deterministic description will generally involve $K+M$ coupled non-linear equations (one for each subclass in the population). We discuss when and how these can be reduced to a smaller set of equations for aggregate quantities. In Sec.~\ref{sec:LNA} we perform then the linear-noise approximation  and use this approximation to characterise the fluctuations about deterministic fixed points. In particular we set up the theory to obtain the spectra of noise-driven quasi-cycles. Using this theory we then present our main results in Sec.~\ref{sec:Consequences}, where we investigate in detail how the heterogeneity in the population affects the properties of stochastic outbreaks of the disease. Finally, in Sec.~\ref{sec:Conclusions} we summarize our findings.

		\section{Model}\label{sec:Model}

We use an extension of the standard SIR model \cite{Kermack1927}, in a population of fixed size $N$. Broadly, each individual can be of one of three types, susceptible (S), infective (I) or recovered (R). The spreading of the disease is described by infection events. These occur either through contact of a susceptible with an infective individual, as described below, or through spontaneous infection. Individuals recover at rate $\rho$, and they die at rate $\kappa$. The death rate is assumed to be independent of the disease status of an individual. To keep the number of individuals in the population constant, any death event is immediately followed by a birth of a new susceptible individual.  This modelling assumption is made for simplicity and is commonly made (see e.g. \cite{Britton2002, Alonso2007, Shulgin1998}).
	
We introduce heterogeneity by dividing the groups of susceptibles and infectives into subclasses. We will write $S_i$ and $I_a$ for these, with $i=1,\dots,K$ and $a=1,\dots, M$.  Individuals in subgroup $S_i$ have susceptibility $\chi_i$ to the disease, and infectives in class $I_a$ have infectiousness $\beta_a$, which describes the propensity of the infective to pass on the disease to susceptible individuals. We write $n_i$ for the number of individuals of type $S_i$, and $m_a$ for the number of individuals in class $I_a$.

	\begin{figure*}[t!]
	\includegraphics[width=0.96\textwidth]{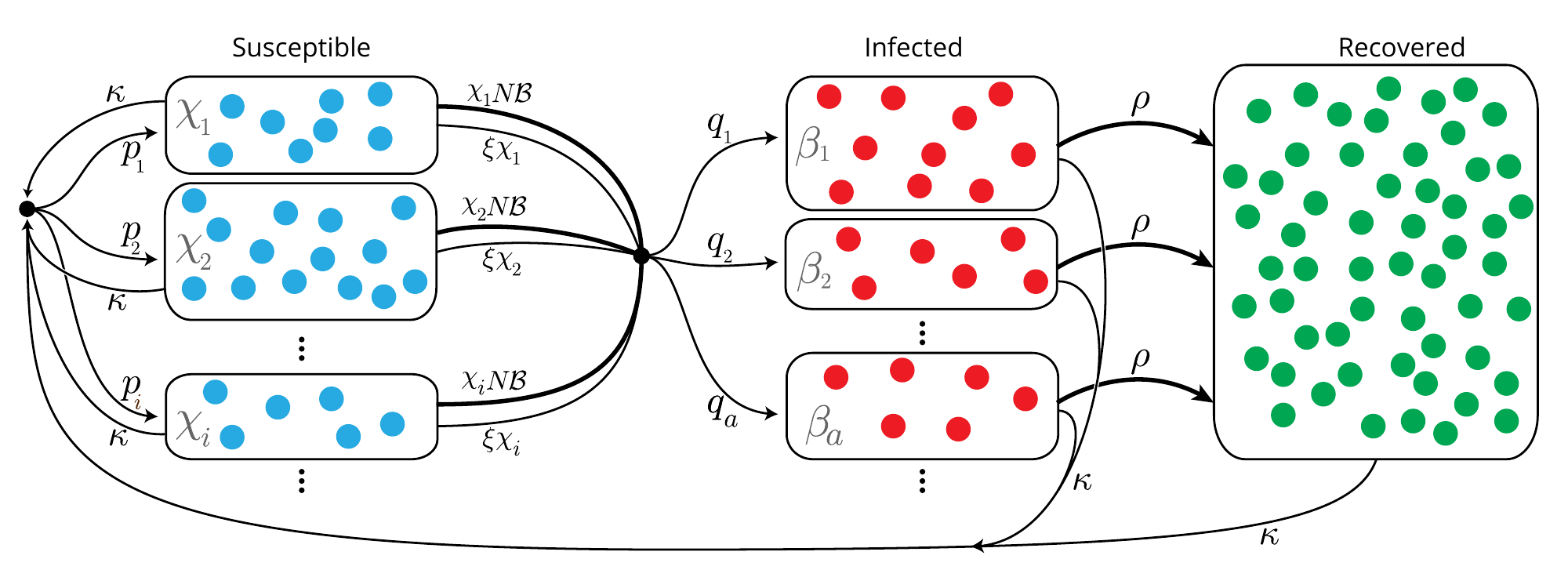}
	\caption{\textbf{SIR model with heterogeneous susceptibility and infectivity.} 
			The diagram illustrates the different processes described by the model. New (susceptible) 
			individuals are born at a rate $\kappa$, and they are assigned a susceptibility of 
			$\chi_{i}$ with probability 
			$p_{i}$. Susceptible individuals transition to an infected state either by spontaneous 
			infection or by contact with any of the infected classes. The former process occurs 
			with rate $\xi\chi_i$, if the susceptible is of type $S_i$. Conact infection occurs 
			at a rate $\chi_i N\fbeta$, where $N\fbeta$ is the total infective power of the 
			population (see Eq.~\eqref{eq:fbeta}). Once infected, the individual is assigned an 
			infectiousness $\beta_{a}$ with probability $q_{a}$. All infected individuals recover 
			at the same rate $\rho$. At any stage, individuals die with a rate $\kappa$.
			To keep the total population $N$ constant, deceased individuals are immediately replaced 
			by a new susceptible individual.
	}
	\label{fig:Model}
	\end{figure*}
	
The dynamics are illustrated in Fig.~\ref{fig:Model}, and can be summarised in the following reaction scheme:
	\BE
	\mbox{Spontaneous infection:~~} & S_{i}\overset{\xi\chi_{i}q_{a}}{\longrightarrow}I_{a} 	\nn \\
	\mbox{Infection by contact:~~}  & S_{i}+I_{a}\overset{\beta_{a}\chi_{i}q_{b}}
															{\longrightarrow}I_{a}+I_{b}		\nn \\
	\mbox{Recovery:~~}				& I_{a}\overset{\rho}{\longrightarrow}R 						\\
	\mbox{Birth/Death:~~} 			& S_{j}\overset{p_{i}\kappa}{\rightarrow}S_{i}				\nn \\
	 								& I_{a}\overset{p_{i}\kappa}{\rightarrow}S_{i}				\nn \\
	 								& R\overset{p_{i}\kappa}{\rightarrow}S_{i},					\nn 
	\label{eq:Model}
	\EE
where $\{p_i\}$ and $\{q_a\}$ represent the probabilities of being assigned a susceptibility $\chi_i$ or infectiousness $\beta_a$ at birth or upon infection, respectively. The first of these reactions describes spontaneous infection, converting an individual in class $S_i$ into an individual of type $I_a$. The per-capita rate of events of this type is $\xi\chi_i q_a$, where $\xi$ is an overall inverse time scale for spontaneous infection, $\chi_i$ is the susceptibility of $S_i$ to the disease, and $q_a$ is the probability that the newly infected individual is in class $I_a$. Similarly, the second reaction describes infection of an individual of type $S_i$ upon contact with an individual of type $I_a$. The newly infected individual is in class $I_b$. Events of this particular type occur with a rate proportional to $\beta_a$ (the propensity of $I_a$ to spread the disease), to $\chi_i$ (the susceptibility of $S_i$) and to $q_b$. The third reaction describes recovery, and the final three reactions are birth/death events. The newly born individual is assumed to be randomly placed into one of the classes $S_i$ ($i=1,\dots,K$), occurring with respective probability $p_i$. We note that our model does not describe potential correlations between the susceptibility of an individual and its infectivity after they become infected; our focus is on heterogeneity of susceptibility due to physiological factors, and not primarily due to contact patterns. Extensions to include correlations can however be constructed among similar lines.

The model defines a continuous-time Markov process, and can be simulated straightforwardly using for example the celebrated Gillespie algorithm \cite{Gillesple1977}. The starting point for the analytical study of the model is the master equation. Our analysis below will be based on approximating the solution to this master equation by performing a system-size expansion \cite{vanKampen1992} and linear-noise approximation, leading to a stochastic differential equation describing the dynamics in the limit of large, but finite population size. 

In order to do this it is useful to first introduce 
	\be
	\overline{\chi}  =  \sum_{i}p_{i}\chi_{i},
	\text{\qquad and \qquad}
	\fchi  =  \frac{1}{N} \sum_{i}\chi_{i}n_{i}.
	\ee
The quantity $\overline\chi$ is the mean susceptibility of a newly born individual, whereas $N\fchi$ describes the aggregate susceptibility of the population. Similarly we define
	\be
	\overline{\beta}    =  \sum_{a}q_{a}\beta_{a}
	\text{\qquad and \qquad}
	\fbeta 	 =  \frac{1}{N} \sum_{a}\beta_{a}m_{a},
	\label{eq:fbeta}
	\ee
where $\overline{\beta}$ represents the mean infectivity of a newly infected individual, and $N\fbeta$ the total `infective power' in the population. We note that $\overline\chi$ and $\overline\beta$ are fixed in time, and are properties of the distributions $\{p_i,\chi_i\}$ and $\{q_a,\beta_i\}$. The quantities $\fchi$ and $\fbeta$, on the other hand, are time-dependent and evolve as the composition of the population changes.

		\section{Deterministic analysis}\label{sec:Deterministic}

			\subsection{Dynamics}

In the limit of an infinite population the dynamics can be described by deterministic equations for the quantities $x_i=\lim_{N\to\infty} n_i/N$, $y_a=\lim_{N\to\infty} m_a/N$. They are given by
	\BE
	\dot{x_{i}} & = & \kappa p_{i}-\kappa x_{i}-\xi\chi_{i}x_{i}-\chi_{i}x_{i}\fbeta,		\nn \\
	\dot{y}_{a} & = & \xi q_{a}\fchi+q_{a}\fchi\fbeta-\rho y_{a}-\kappa y_{a}.
	\label{eq:Deterministic_Si,Ia-phi,psi}
	\EE
These ordinary differential equations can be derived either by using direct mass-action kinetics, or from the lowest-order expressions in an expansion of the master equation in the inverse system size \cite{vanKampen1992}.

Ultimately we will mostly be interested in aggregate quantities, i.e. the total density of susceptibles or infectives in the population, irrespective of  what subclass they belong to. We therefore introduce
	\be
	S=\sum_i x_i \text{\qquad and \qquad} I=\sum_a y_a.
 	\ee
From Eqs.~\eqref{eq:Deterministic_Si,Ia-phi,psi} we find
	\BE
	\dot{S} & = & \kappa-\kappa S-\xi\fchi-\fbeta\fchi,		\nn \\
	\dot{I} & = & \xi\fchi+\fchi\fbeta-\rho I-\kappa I.
	\label{eq:det1}
	\EE
This system is not closed due to the presence of $\fchi$ and $\fbeta$ on the right-hand side. These quantities in turn evolve in time according to
	\BE
	\dot{\fchi}	 & = & \kappa\overline{\chi}-\kappa\fchi-(\xi+\fbeta)\sum_{i}\chi_{i}^{2}x_{i},	\nn \\
	\dot{\fbeta} & = & \xi\fchi\overline{\beta}+\overline{\beta}\fchi\fbeta-(\rho+\kappa)\fbeta,	
	\label{eq:det2}
	\EE
which again does not close the set of equations, due to the presence of the term $\fchi_2(t)\equiv\sum_{i}\chi_{i}^{2}x_{i}(t)$. Modulo normalisation and recalling that the $\{x_i\}$ are time-dependent, this object is recognised as the second moment of the distribution of susceptibilities among the group of susceptibles {\em at time $t$}. It cannot be determined from Eqs.~\eqref{eq:det1} and \eqref{eq:det2} alone. Instead we find
	\BE
	\dot{\fchi}_{n}=\kappa\overline{\chi^{n}}-\kappa\fchi_{n}-(\xi+\fbeta)\fchi_{n+1},
	\label{eq:phin}
	\EE
where we have introduced $\overline{\chi^{n}}=\sum_{i}p_{i}\chi_{i}^{n}$ and $\fchi_{n}=\sum_{i}x_{i}\chi_{i}^{n}$. This indicates that the deterministic dynamics at the aggregate level is described by an infinite hierarchy of equations. This set of equations does not close in the transient regime. However, as we will see next, closure can be achieved assuming the system settles down to a fixed point in the long run.  

			\subsection{Fixed point}\label{sub:Fixed-point}

We proceed by a brief analysis of the fixed points of the deterministic dynamics. We will label these by a star. They can be obtained by setting $\dot{x}_i=0$ and $\dot{y}_a=0$ in Eqs.~\eqref{eq:Deterministic_Si,Ia-phi,psi}, leading to
	\BE
	x_{i}^{\star} & = & \frac{\kappa p_{i}}{\kappa+\left(\xi+\fbeta^{\star}\right)\chi_{i}},	\nn \\
	y_{a}^{\star} & = & \frac{\left(\xi+\fbeta^{\star}\right)\fchi^{\star}q_{a}}{\rho+\kappa}.
	\label{eq:FixedPoint_Si,Ia}
	\EE
Similarly, we find the fixed points of the aggregate quantities $S$, $I$, $\fchi$ and $\fbeta$ from Eqs.~(\ref{eq:det1},\ref{eq:det2}). After re-arranging and using Eqs.~\eqref{eq:FixedPoint_Si,Ia} we arrive at
	\BE
 	S^{\star} 	  & = & 1-\frac{\left(\rho+\kappa\right)}{\kappa}\frac{\fbeta^{\star}}
 																		{\overline{\beta}},		\nn \\
	I^{\star} 	  & = & \frac{\fbeta^{\star}}{\overline{\beta}},								\nn \\
	\fchi^{\star} & = & \frac{\left(\rho+\kappa\right)}{\left(\xi+\fbeta^{\star}\right)}
												\frac{\fbeta^{\star}}{\overline{\beta}},		\nn \\
	\fbeta^{\star}& = & \frac{\overline{\beta}\kappa}{\left(\rho+\kappa\right)}\sum_{i}\left(
					\frac{\chi_{i}p_{i}}{\frac{\kappa}{\xi+\fbeta^{\star}}+\chi_{i}}\right).
	\label{eq:FixedPoints-Closed}
	\EE
which is a closed set of equations, for a given set of model parameters $\{p_i,\chi_i,q_a,\beta_a\}$. 

We highlight that while the transient dynamics of the system described in terms of the four macroscopic variables $S$, $I$, $\fchi$ and $\fbeta$ generates an infinite hierarchy of equations, potential fixed points can be uniquely described by a closed set of equations, assuming that the distribution of susceptibilities at birth and of the propensity of newly infected individuals to pass on the disease are known. In other words, the fixed point can be obtained in terms of the model parameters $\{q_a, \beta_a\}$ and $\{p_i, \chi_i\}$. While we cannot provide an analytical proof that the deterministic system will always converge to a fixed point, we note that, for the range of parameter used, we have not detected a single case in which numerically integrating Eqs. (\ref{eq:Deterministic_Si,Ia-phi,psi}) did not lead to a fixed point. In this context it is useful to point out that, in a homogeneous model, any combination of susceptibility and infectivity within the range of parameters used here would lead to a basic reproductive number above unity. For such models it is known that stable fixed points are eventually reached \cite{Keeling2008}.

		\section{Linear-noise approximation}\label{sec:LNA}

We now proceed to analyse the effects of stochasticity in the model, with a particular focus on the interaction between heterogeneity of individuals in the population and the noise induced by the demographics of the finite system.

\begin{figure*}[!ht]
	\includegraphics[width=0.96\textwidth]{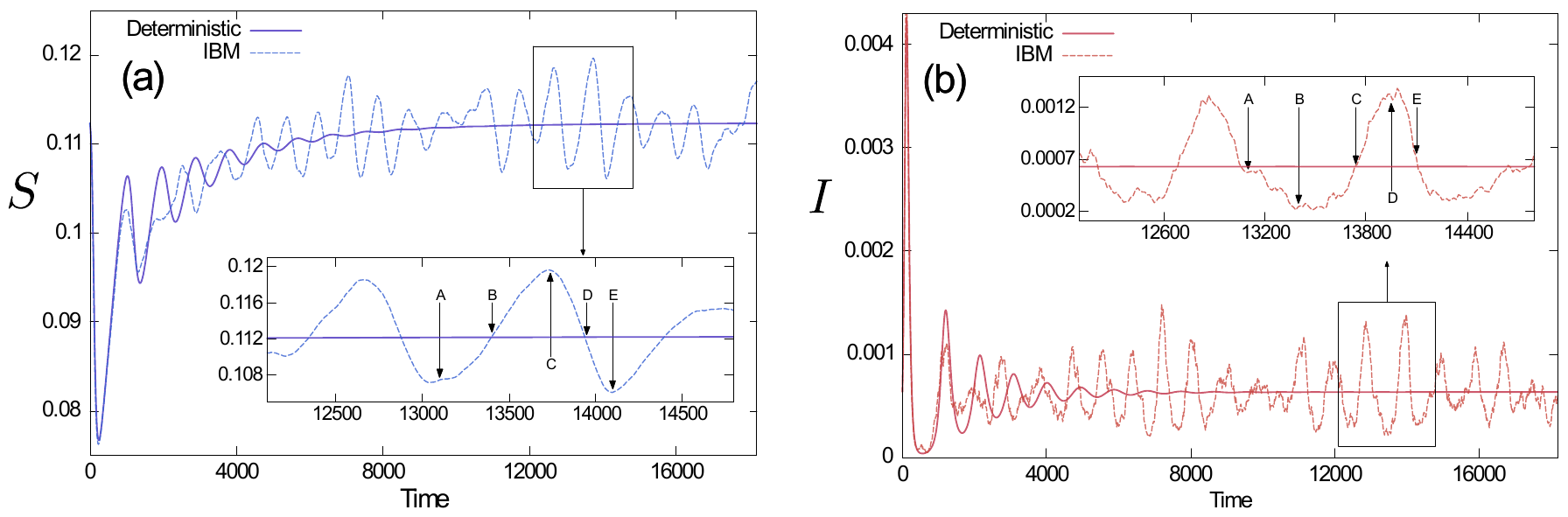}
	\caption{\textbf{Population dynamics.} Time series of the population density 
			of total susceptible (panel  \textbf{(a)}) and total infected individuals 
			(panel  \textbf{(b)}). Noise-sustained oscillations are clearly seen. The 
			insets show a zoom in on the cycles. Labels $A, B,\dots, E$ are for later 
			purposes (see below).}
	\label{fig:PopulationDynamics}
	\end{figure*}

We illustrate these effects in Fig.~\ref{fig:PopulationDynamics}, and show an example of both the deterministic time-evolution of the system (thick continuous lines) and a realization of an individual-based simulation (thin dashed lines); the latter illustrates the intrinsic stochasticity of the process. Even after the deterministic model has reached a fixed point, the individual-based model shows sustained oscillations around it. We will focus our attention on these stochasticity-driven periodic outbreaks in the remainder of this article. In particular we will study how the heterogeneity in the population affects their properties.

			\subsection{Stochastic Dynamics}\label{sub:stoch}

In order to carry out an analysis of the stochastic dynamics, we write $n_i/N=x_i+\tilde x_i/\sqrt{N}$, and $m_a/N=y_a+\tilde y_a/\sqrt{N}$, where $x_i(t)$ and $y_a(t)$ are the solutions of the deterministic equations (\ref{eq:Deterministic_Si,Ia-phi,psi}) and the quantities with a tilde describe the stochastic fluctuations about the deterministic trajectory. The above ansatz reflects the anticipation that these fluctuatons will have a relative magnitude of order $N^{-1/2}$. We then carry out an expansion in the inverse system size up to and including sub-leading order \cite{vanKampen1992}, and arrive at
	\BE
	\label{eq:LNA_Si,Ia}
	\dot{\tilde x}_i	& = & -\kappa\tilde{x}_i-\left(\xi+\fbeta^{\star}\right)
						\chi_{i}\tilde{x}_{i}-\chi_{i}x_{i}^{\star}\tilde{\fbeta}+\eta_{i},		\nn \\
	\dot{\tilde y}_a 	& = & q_a\left(\xi\tilde{\fchi}+\tilde{\fchi}\fbeta^{\star}+\fchi^{\star}
				\tilde{\fbeta}\right)-\left(\rho+\kappa\right)\tilde{y}_{a}+\nu_{a}.	
	\EE
The $\{\eta_{i}\}$ and $\{\nu_{a}\}$ are Gaussian white noise variables, with variance and co-variance (across components) as described in more detail in the Supplement (see \ref{app:Noise-correlators}). Writing $\tilde S=\sum_i \tilde x_i$ and $\tilde I=\sum_a \tilde y_a$ we find the following dynamics of fluctuations at the aggregate level, 
	\BE
	\label{eq:LNA}
	\dot{\tilde S}		 & = & -\kappa\tilde{S}-\left(\xi+\fbeta^{\star}\right)\tilde{\fchi}-
											\fchi^{\star}\tilde{\fbeta}+\sum_{i}\eta_{i},		\nn \\
	\dot{\tilde I} 		 & = & \left(\xi+\fbeta^{\star}\right)\tilde{\fchi}+\fchi^{\star}
									\tilde{\fbeta}-(\rho+\kappa)\tilde{I}+\sum_{a}\nu_{a},		\nn \\
	\dot{\tilde{\fchi}}	 & = & -\kappa\tilde{\fchi}-\fchi_{2}^{\star}\tilde{\fbeta}-
									\left(\xi+\fbeta^{\star}\right)\sum_{i}\chi_{i}^{2}
										\tilde{x_{i}}+\sum_{i}\chi_{i}\eta_{i},					\nn \\
	\dot{\tilde{\fbeta}} & = & \left(\xi+\fbeta^{\star}\right)\overline{\beta}\tilde{\fchi}+
									\overline{\beta}\fchi^{\star}\tilde{\fbeta}-\left(\rho+\kappa
											\right)\tilde{\fbeta}+\sum_{a}\beta_{a}\nu_{a}.		
	\EE
As in the deterministic analysis, this set of equations for the transient dynamics is not closed. This is due to the term $\sum_{i}\chi_{i}^{2}\tilde{x_{i}}$ in the equation for $\dot{\tilde{\fchi}}$. However, as in Section~\ref{sub:Fixed-point}, we will show below that a closed set of equations for fluctuations in the stationary state can be derived.

			\subsection{Fluctuation around the deterministic fixed point}\label{sub:Fluctuations}

We here show that although Eqs.~\eqref{eq:LNA} are not closed, we can explore noise-induced oscillations around the deterministic fixed point. To this end we introduce the Fourier transforms (with respect to time) of the variables $\tilde x_i$  and $\tilde y_a$. We will denote these by $\widehat{x}_i$ and $\widehat{y}_a$. From the Langevin equations (\ref{eq:LNA_Si,Ia}) we find, after re-arranging,
	\BE
	\label{eq:Fourier_Si,Ia_re}
	\widehat{x}_i & = & \frac{-\chi_{i}x_{i}^{\star}\widehat{\fbeta}+
							\widehat{\eta}_i}{i\omega+\kappa+\left(\xi+
								\fbeta^{\star}\right)\chi_{i}},							\nn \\
	\widehat{y}_a & = & \frac{\left[\left(\xi+\fbeta^{\star}\right)
							\widehat{\fchi}+\fchi^{\star}\widehat{\fbeta}\right]
							q_{a}+\widehat{\nu}_a}{i\omega+\rho+\kappa}.
	\EE
The noise variables $\{\eta_i\}$ and $\{\nu_a\}$ are uncorrelated in time, and their variance and correlation across components can be expressed in terms of known quantities (see Eqs.~\eqref{eq:noisecorr} in the Supplement). The variable $\omega$ is the conjugate of time under Fourier transform. Similarly, we find the following for the relevant aggregate quantities,
	\BE
	\label{eq:Fourier-AiCDE}
	\widehat{S} & = & \frac{1}{i\omega+\kappa}\left[-
						\frac{i\omega+D}{\overline{\beta}}\widehat{\fbeta}+\frac{1}
							{\overline{\beta}}\sum_{a}\beta_{a}\widehat{\nu}_a+\sum_{i}
								\widehat{\eta}_i\right],										\nn \\
	\widehat{I} & = & \frac{1}{i\omega+D}\left[
						\frac{i\omega+D}{\overline{\beta}}\widehat{\fbeta}-
							\frac{1}{\overline{\beta}}\sum_{a}\beta_{a}\widehat{\nu}_a+
								\sum_{a}\widehat{\nu}_a\right],									\nn \\
	\widehat{\fchi}  & = & \frac{1}{\overline{\beta}C}\left[\left(i\omega+E\right)
							\widehat{\fbeta}-\sum_{a}\beta_{a}\widehat{\nu}_a\right],			\nn \\
	\widehat{\fbeta} & = & \frac{\overline{\beta}C\sum\limits _{i}\frac{\chi_{i}
								\widehat{\eta}_i}{i\omega+A_{i}}+
								\sum\limits _{a}\beta_{a}\widehat{\nu}_a}{
								i\omega+E+\overline{\beta}C\kappa\sum\limits _{i}
								\frac{\chi_{i}^{2}p_{i}}{A_{i}\left(i\omega+A_{i}\right)}},
	\EE
where, for simplicity, we have introduced the notation 
	\BE
	\label{eq:AiCDE}
	A_{i} 	& = & \kappa+\left(\xi+\fbeta^{\star}\right)\chi_{i},				\nn \\
	C 		& = & \xi+\fbeta^{\star},											\nn \\
	D 		& = & \rho+\kappa,													\nn \\
	E 		& = & \rho+\kappa-\overline{\beta}\fchi^{\star}.
	\EE
Eqs.~\eqref{eq:Fourier-AiCDE} constitute a closed set of equations for the Fourier transforms of the aggregate fluctuations $\tilde S, \tilde I, \tilde \fchi$ and $\tilde \fbeta$  in the stationary state. We thus make an observation similar to that in Section~\ref{sec:Deterministic}: although we cannot describe the evolution of fluctuations in the transient regime, we can derive a closed description of the statistics of fluctuations about deterministic fixed points within the linear-noise approximation. 

			\subsection{Power Spectral Density}\label{sub:PSD}

Eqs.~\eqref{eq:Fourier-AiCDE} can be used describe the periodic cycles shown in Fig.~\ref{fig:PopulationDynamics}; we will now proceed to analyse these in more detail. Specifically we will use the above results to compute the power spectral density (PSD) of fluctuations. This allows us to identify the characteristic frequency of noise-driven epidemic cycles, and to infer information about their amplitude.  

The (average) power spectral density of a time series, $z(t)$, generated from the stochastic individual-based model is given by $\mathcal{P}_{z}(\omega)=\langle|\widehat{z}(\omega)|^{2}\rangle$, where $\avg{\cdots}$ stands for an average over realizations of the stochastic dynamics. The PSD can be computed analytically for all individual signals $x_i$, $y_a$, and for the aggregate variables $S$, $I$, $\fchi$ and $\fbeta$. The resulting expressions are lengthy; for completeness we provide them in the Supplement (see \ref{app:PSDs}).  As an illustration we here show the PSD of $\fbeta$,
	\BE
	\label{eq:PSD_psi}
	\mathcal{P}_{\fbeta} (\omega)& = & \frac{2\fchi^{\star}C}{|g|^{2}}\left(\overline{\beta^{2}}-
								\frac{\overline{\beta}^{2}C\kappa}{D}
								\sum\limits _{i}\frac{\chi_{i}p_{i}A_{i}}{
								\omega^{2}+A_{i}^{2}}\right)							
								 -\frac{\left(\overline{\beta}C\kappa\right)^{2}}{|g|^{2}}
								 \left[\sum\limits _{i,j}\frac{p_i p_j \chi_i \chi_j (A_i+A_j)
								 (\omega^2+A_iA_j)}{A_i A_j (\omega^2+A_i^2)(\omega^2+A_j^2) }\right],	
	\EE
with
	\be
	\label{eq:gg*}
	|g|^{2} = \left[E+\overline{\beta}C\kappa\sum\limits _{i}\frac{\chi_{i}^{2}p_{i}}
				{\omega^{2}+A_{i}^{2}}\right]^{2}+\omega^{2}\left[1-
				\overline{\beta}C\kappa\sum\limits _{i}\frac{\chi_{i}^{2}p_{i}}{A_{i}
				\left(\omega^{2}+A_{i}^{2}\right)}\right]^{2}.
	\ee
 As detailed in the Supplement (see Sec. \ref{app:PSDs}) the power spectra of $S, I$ and $\fchi$ can be expressed in terms of that of $\fbeta$; many of the characteristics of the spectra of $S, I$ and $\fchi$ are shared with those of $\fbeta$, or directly related to it. We note that the RHS of Eq.~\eqref{eq:PSD_psi} is proportional to $1/|g|^2$, and the same is the case for the spectral densities of $\fchi, S$ and $I$ (see Eqs.~\eqref{eq:PSD}); as a result, some of the key properties of the power spectra are determined by the behaviour of $|g|^2$, as discussed in more detail below.

	\subsection{Test Against Simulations}
\begin{figure*}[t!!]
	\includegraphics[width=0.96\textwidth]{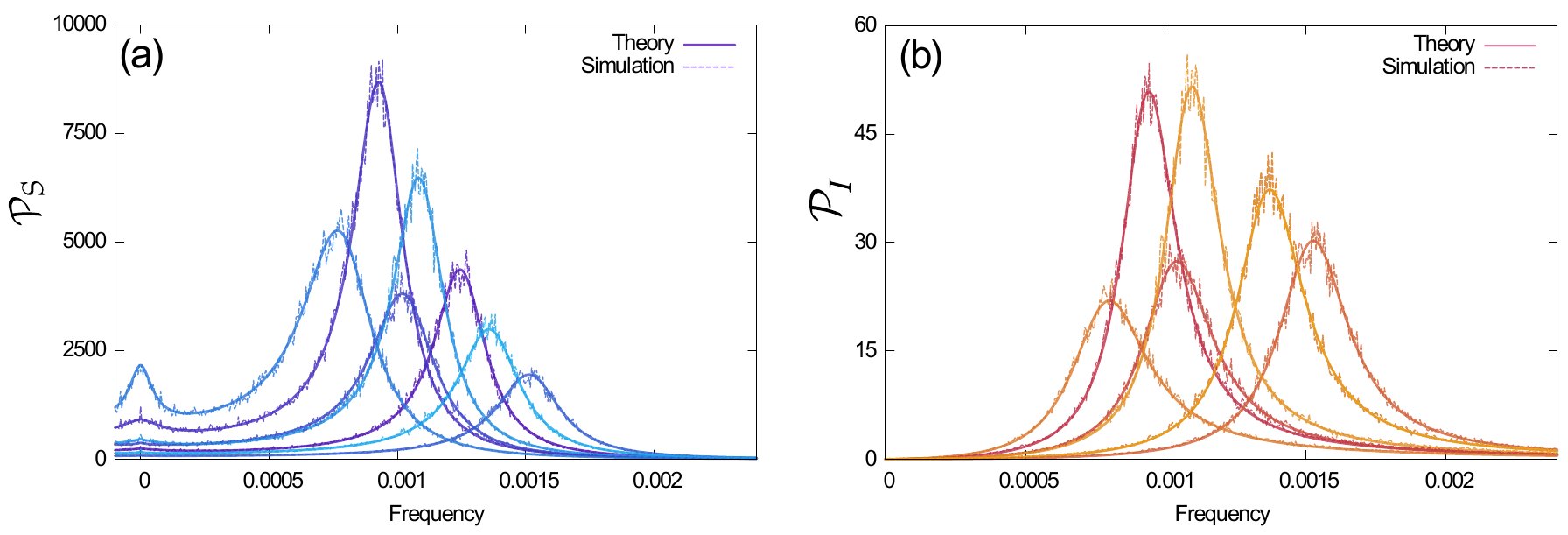}
	\caption{\textbf{Power spectral densities} of the fluctuations of \textbf{(a)} 
			Susceptible and \textbf{(b)} 
			Infected population for seven different examples of the model, generated as 
			explained in more detail in the text. In all cases theory and simulations 
			agree.}
	\label{fig:PSD}
	\end{figure*}

To illustrate the model and test our analytical results, we sampled possible heterogeneous populations. Specifically, the simulations shown in Fig.~\ref{fig:PSD} are for populations with five susceptible and three  infected subclasses. For each example, the probabilities $\{p_i\}$ and $\{q_a\}$ were drawn at random from a flat distribution over the simplexes $\sum_{i}p_{i}=1$ and $\sum_{a}q_{a}=1$. Susceptibilities and infectivities were assigned randomly in the intervals $0.5\leq\chi_{i}\leq2.5$ and $0.3\leq\beta_{a}\leq1.3$. Simulations are for $N=10^{6}$, and the rates for recovery, birth/death and immigration were set at $\rho=0.07~$, $\kappa=5.5\times10^{-5}$ and $\xi=5\times10^{-6}$ respectively. The rates $\beta_a$, $\rho$, $\kappa$ and $\xi$ have units of $\mbox{days}^{-1}$, whereas $\chi_i$ is dimensionless. The chosen rates are representative of childhood diseases such as whooping cough, measles, rubella or chickenpox \cite{Anderson1992}.

The resulting PSDs are shown in Fig.~\ref{fig:PSD}. The continuous thick lines show the analytical result, and dashed lines are obtained from simulations, as an average over realizations of the individual-based model. As can be seen from the figure, the predictions of Eqs.~\eqref{eq:PSD} precisely match the results from simulations. In all figures, axes labelled `frequency' show $f=\omega/2\pi$, and have units of $\mbox{days}^{-1}$.

		\section{Consequences of Heterogeneity}\label{sec:Consequences}

Having established an analytical description of quasi-cycles, we now use this theory to identify which properties of the distribution of $p_i$, $\chi_i$, $q_a$ and $\beta_a$ are most relevant for the characteristics of stochastic quasi-cycles in heterogeneous populations. Specifically, we study how heterogeneity in the population affects the dominant frequency of quasi-cycles, their amplitude and the sharpness of the spectra. We will then also discuss if and how the different subgroups synchronise during the epidemic cycles.

			\subsection{Dominant Cycle Frequency}\label{sub:Freq}

Numerical inspection of the different terms in the analytical solution of the PSDs suggests that the dominating element is the factor $1/|g|^{2}$, as briefly indicated in Sec.~\ref{sub:PSD}. The frequency for which $|g|^{2}$ reaches its minimum roughly corresponds to the dominant cycle frequency, $\omega_d$, in the PSDs. The minimum of $|g|^{2}$ can be found by differentiation of the expression in Eq.~\eqref{eq:gg*}.  Assuming that $C\chi_{i}\gg\kappa$ we further approximate the location of this minimum. This assumption is valid if infection processes occur on a time scale which is much shorter than the life expectancy of an individual. Further, we assume that $\omega\gg A_{i}$, i.e. that a  susceptible individual typically lives through several epidemic events before it becomes infected. Both approximations are intuitively plausible for childhood diseases, known to show periodic outbreaks \cite{Anderson1992}. Making these assumptions we find that the frequency for which $|g|^{2}$ is minimal can be approximated as
	\be
	\omega_{d}\approx\sqrt{\kappa\overline{\chi}\overline{\beta}}.
	\label{eq:Approx-Frequency}
	\ee
This implies that the characteristic frequency is determined (mostly) by the mean susceptibility at birth and the mean infectivity at infection ($\overline{\chi}$ and $\overline{\beta}$) and the capacity of replenishment of the susceptible pool ($\kappa$). 

\begin{figure*}[t!!]
	\includegraphics[width=0.96\textwidth]{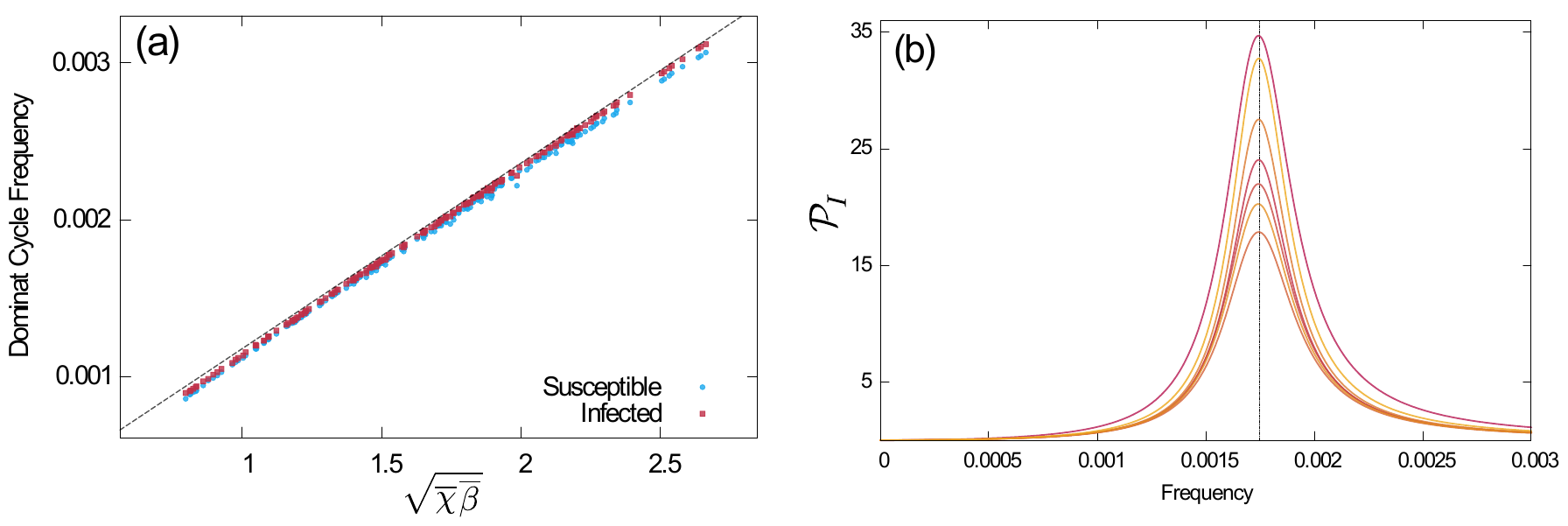}
	\caption{\textbf{Verification of approximation \eqref{eq:Approx-Frequency} for the dominating 
				frequency of cycles.} \textbf{(a)} Frequency $f=\omega/2\pi$ at the maximum of 
				the PSD, determined from  Eqs.~\eqref{eq:PSD} as a function of 
				$\sqrt{\overline\chi \overline\beta}$, for fixed $\kappa$. The black dashed line 
				corresponds to Eq.~\eqref{eq:Approx-Frequency}.
				Markers are from $200$ different populations, each with $5$ susceptible 
				and $3$ infected subgroups, and with random choices of 
				$\{p_i,\chi_i,q_a,\beta_a\}$. The values of $\chi_{i}$ and $\beta_{a}$ 
				were chosen from the interval $1.7\pm1.6999995$; $q_a$ and $p_i$ 
				from a flat distribution. This resulted in 
				values of $\overline{\chi}$ and $\overline{\beta}$ in the range 0.3 to 3.3, and for 
				$\overline{\chi^{2}}$ and $\overline{\beta^{2}}$ in the range 0.1 to 10.
				\textbf{(b)} PSD of the total infected population of different random distributions 
				of $\{p_i,\chi_i,q_a,\beta_a\}$, with equal values for $\overline{\chi}$ 
				and $\overline{\beta}$, but different values of $\overline{\chi^{2}}$ and 
				$\overline{\beta^{2}}$. As a consequence of Eqs.~\eqref{eq:Approx-Frequency} 
				and \eqref{eq:Approx-Amplitude}, the characteristic frequency is the same for all 
				such samples, but the height of the peak in the PSD varies significantly.
				The vertical dotted line is a visual aid.}
	\label{fig:Frequency}
	\end{figure*}

The validity of our approach is confirmed in Fig.~\ref{fig:Frequency}(a), where we test the approximation against simulations for a wide set of parameters. A perhaps more intuitive representation of our result can be found in  Fig.~\ref{fig:Frequency}(b), where we show the power spectra of several sample populations, each with different distributions of $\{p_i,\chi_i,q_a,\beta_a\}$, but all with the same first moments $\overline{\chi}$ and $\overline{\beta}$. As seen in the figure, this produces spectra of different amplitudes but with the same characteristic frequency.

			\subsection{Amplitude of Stochastic Cycles}

While we have found above that the dominant frequency of stochastic cycles is largely determined by the first moments $\overline{\chi}$ and $\overline{\beta}$, the results shown in Fig.~\ref{fig:Frequency}(b) demonstrate that this is not the case for the amplitude of the spectra at the dominant frequency. To investigate this further we evaluate the analytic expressions for the PSDs in Eqs.~\eqref{eq:PSD} at the approximation of $\omega_d$ in Eq.~\eqref{eq:Approx-Frequency}. Making the same assumptions as in Section~\ref{sub:Freq}, we find that the height of the peak in the power spectra can be approximated as
	\BE
	\mathcal{P}_{I}\left(\omega_d \right)
				 & \approx & \frac{2\left(\rho+\kappa\right)}{
							\left[\frac{\left(\rho+\kappa\right)\xi}{
							{\fbeta^{\star}}}+\frac{\fbeta^{\star}\overline{\chi^{2}}}
							{\overline{\chi}}\right]^{2}}\frac{\overline{\beta^{2}}}
							{\overline{\beta}^{3}},										\nn \\
	\mathcal{P}_{S}\left(\omega_d \right)
				 & \approx &  \frac{\left(\rho+\kappa\right)^{2}}{\kappa
							\overline{\chi}\overline{\beta}} \mathcal{P}_{I}\left(\omega_d \right).
	\label{eq:Approx-Amplitude}
	\EE
We note the presence of the second moments $\overline{\chi^{2}}$ and $\overline{\beta^{2}}$, unlike in Eq.~\eqref{eq:Approx-Frequency}. This indicates that the spread of susceptibilities and infectivities is relevant to the size of the epidemic.

\begin{figure*}[t!!]
	\includegraphics[width=0.96\textwidth]{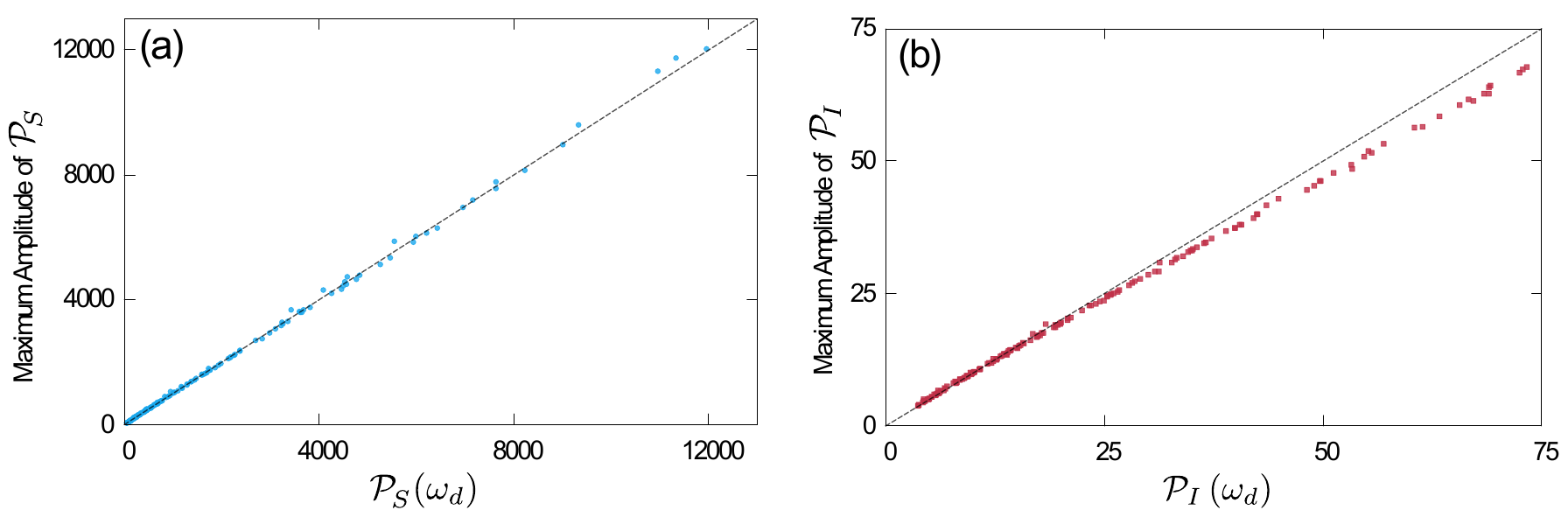}
	\caption{\textbf{Verification of approximation (\ref{eq:Approx-Amplitude}) for the peak-height 
				of the spectral densities.} Horizontal axes show the prediction 
				of Eqs.~\eqref{eq:Approx-Amplitude} for susceptibles \textbf{(a)}, and infectives 
				\textbf{(b)}. On the vertical axis we show the height at the peak of the spectra, 
				as determined numerically from Eqs.~(\ref{eq:PSD}).  Black dashed lines are the 
				diagonal (`$y=x$'), and markers represent the populations described in 
				Fig.~\ref{fig:Frequency}.}
	\label{fig:Amplitudes}
	\end{figure*}

In Fig.~\ref{fig:Amplitudes} we plot results from the approximation in Eqs.~\eqref{eq:Approx-Amplitude} against the maximum amplitude of spectra obtained numerically from the full expression (within the LNA), see Eqs.~\eqref{eq:PSD} in the Supplement. The data confirms that the approximation is valid for a wide range of parameters. While we find slight deviations at large amplitudes in the case of the infectives, the approximation is very robust for the susceptible population.

			\subsection{Sharpness of the Spectra}
			
We now turn to the sharpness of the peak in the PSDs. The sharper the peak, the closer the stochastic outbreaks are to perfect cyclic behaviour. Conversely, cyclic behaviour is less distinct if the peak in the spectrum is shallow. This has been described before as the `coherence' of the spectra \cite{Alonso2007}. As we will investigate a different notion of coherence in Sec.~\ref{sub:Synchro}  and in order to avoid confusion, we will refer to the concentration of power near the peak of the spectrum as `sharpness'.

Following \cite{Alonso2007}, we define the sharpness as the relative spectral power accumulated in an interval around the peak,
	\be\label{eq:s}
	\mathbb{S}=\frac{\int\limits _{\omega_d-\Delta\omega}^{\omega_d+\Delta\omega}
			\mathcal{P}(\omega)~d\omega}{\int\limits _{-\infty}^{+\infty}\mathcal{P}(\omega)~d\omega}.
	\ee
We compute the sharpness  numerically, using the expressions in Eqs.~\eqref{eq:PSD}. In order to evaluate the denominator in Eq.~\eqref{eq:s} we integrate up to an upper cutoff of $\omega_{max}=\pi/100~\mbox{days}^{-1}$. In the numerator we use $\Delta\omega=0.05\,\omega_{max}$. The choice of $\Delta\omega$ can be illustrated using Fig.~\ref{fig:Frequency}(b), where the sharpness $\mathbb{S}$ of the peak roughly corresponds to the fraction of total power concentrated in the interval between frequencies of $0.0015$ and $0.002$ days$^{-1}$.

\begin{figure*}[t!!]
	\begin{center}
	\includegraphics[width=0.48\textwidth]{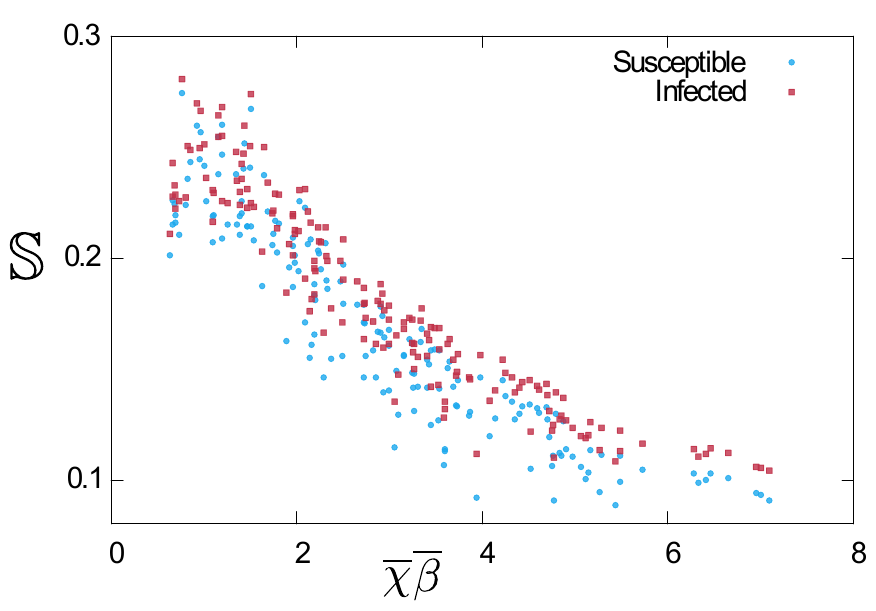}
	\end{center}
	\caption{\textbf{Sharpness of the power spectra} as a function of the product of 
				the mean susceptibilities and infectivities at birth/infection. Data is for 
				the populations described in Fig.~\ref{fig:Frequency}}
	\label{fig:Sharpness}
	\end{figure*}

In Fig.~\ref{fig:Sharpness} we show the sharpness of spectra for $200$ random populations (as described in Fig~\ref{fig:Frequency}). It is clear from the figure that there is a trend of increasing sharpness as the product of the mean susceptibility and infectivity at birth approaches unity (in the dimensions used here). The spread of the markers on the vertical axis indicates that there are significant effects of heterogeneity. It proves difficult, though, to find a functional dependence on higher moments of the distributions of susceptibilities and/or infectivities which would further collapse the data.

			\subsection{Synchronization between Subgroups}\label{sub:Synchro}

\begin{figure*}[t!!]
	\includegraphics[width=0.96\textwidth]{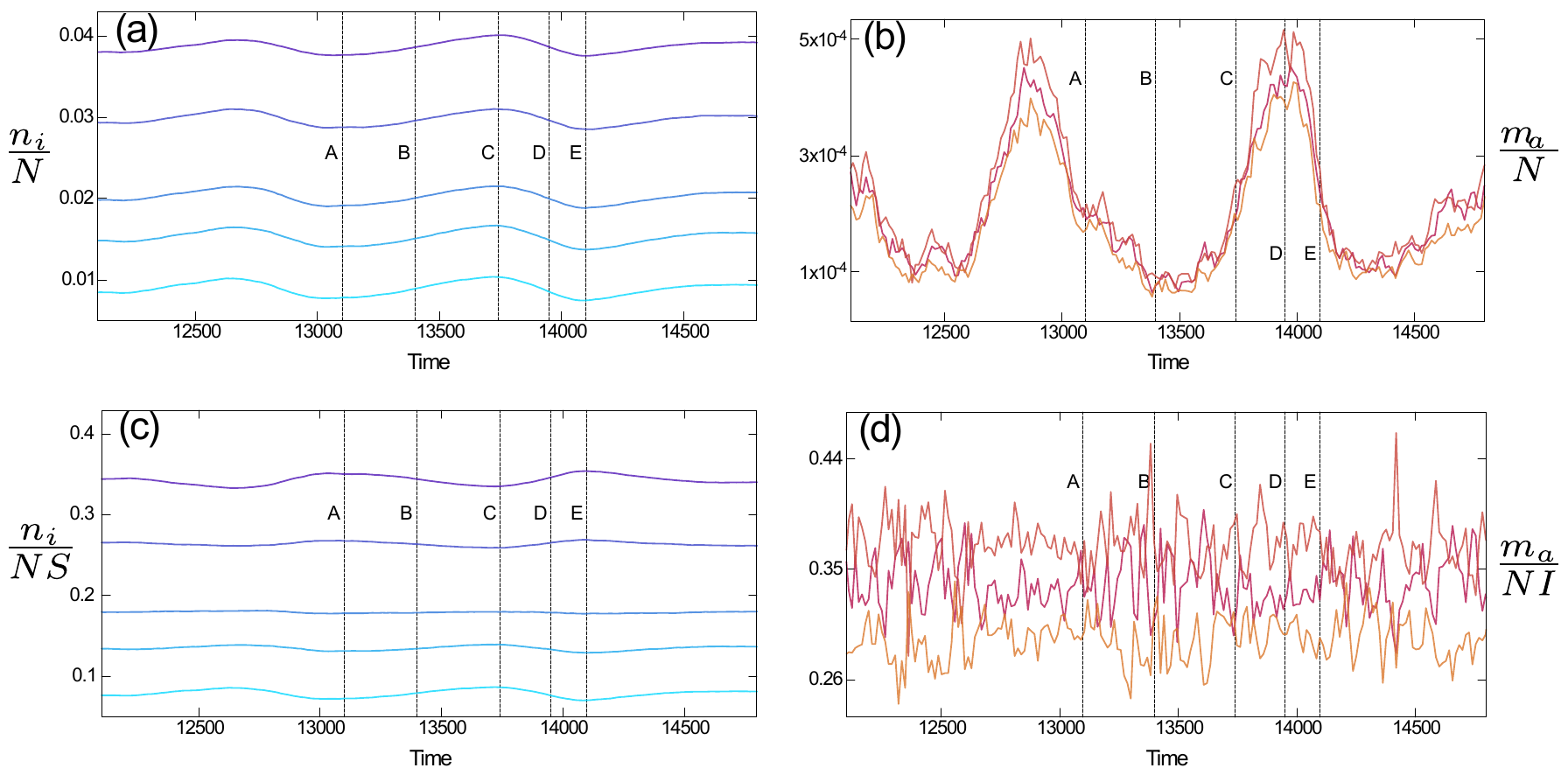}
	\caption{\textbf{Stochastic cycles in subgroups of susceptibles and infectives.} We show the 
				same simulation run as in Fig.~\ref{fig:PopulationDynamics}, but now split up into 
				the different subgroups. Panels \textbf{(a)} and \textbf{(b)} show the number of 
				individuals in each susceptible and infective subgroup normalised by the total 
				population ($N$). In panels \textbf{(c)} and \textbf{(d)}, we show the number of 
				individuals in each subgroup divided by the total number of susceptible or infected 
				individuals, respectively ($NS$ and $NI$). Lines labelled $A$ to $E$ refer to points in the 
				cycles of the aggregate variables $S$, $I$ shown in Fig.~\ref{fig:PopulationDynamics}.} 
	\label{fig:SubgroupCycles}
	\end{figure*}

We have established so far that introducing heterogeneity leads to significant changes in the quasi-cycles of the aggregate numbers of susceptible and infective individuals. However, we have not yet said much about the dynamics of the individual subgroups. In Fig.~\ref{fig:SubgroupCycles} we show the same example of sustained oscillations as in the inset of Fig.~\ref{fig:PopulationDynamics}, but instead of the total susceptible and infected population we now highlight the time evolution of each of the subgroups. 

\begin{figure*}[t!!]
	\includegraphics[width=0.96\textwidth]{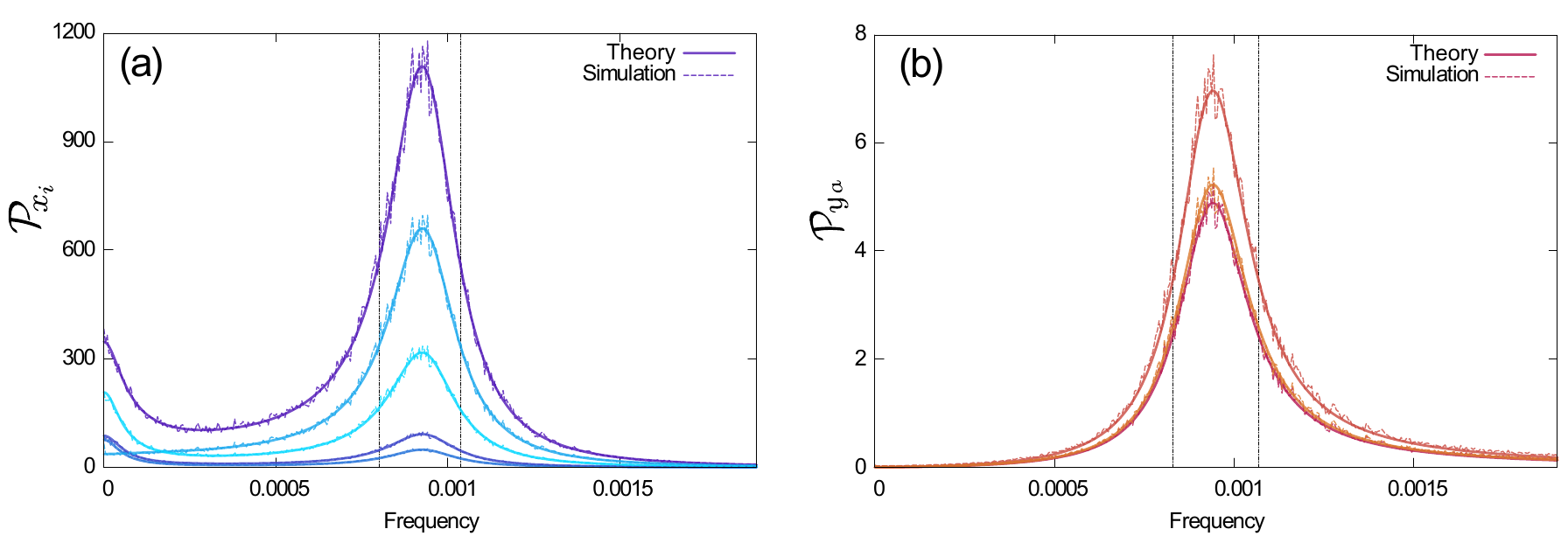}
	\caption{\textbf{Power spectra of fluctuations for different subclasses of susceptibles and 
				infectives}. We use the same sample of the model parameters 
				$\{\chi_i, p_i, \beta_a, q_a\}$ as in Fig.~\ref{fig:PSD}. Simulations 
				are averaged over multiple realizations of the stochastic dynamics, at fixed model 
				parameters. The vertical dotted lines are for later purposes and mark the locations at 
				which the power spectra take values approximately equal to half the maximum amplitude.}
	\label{fig:PSD_Subclasses}
	\end{figure*}

In the upper two panels, (a) and (b), we show time series of the number of individuals in each subgroup normalised by the total population size. More specifically, we show susceptible subclasses ($n_i/N$) in panel (a), and infective subclasses ($m_a/N$) in panel (b). For each of these, stochastic oscillations can be observed. These cycles are pronounced for the case of the infective subgroups, panel (b), and more shallow for the susceptibles, panel (a). This is to be expected, given that the total number of susceptibles is more than an order of magnitude larger than those of the infectives (see also Fig.~\ref{fig:PopulationDynamics}).  From Fig.~\ref{fig:SubgroupCycles} (a) and (b) it is clear that all subgroups undergo cycling of roughly the same frequency. This is confirmed by the power spectra in Fig.~\ref{fig:PSD_Subclasses}.

We note that these statements rely on expressing number of individuals in each class as a fraction of the total population, and not relative to the time-dependent total number of susceptibles or infectives respectively. We contrast the above with a representation in which we express the occupancy in each infective subgroup as a fraction of the infectives only, and similarly for the susceptibles. To this end we replot the simulation run shown in Fig.~\ref{fig:SubgroupCycles} (a) and (b), but now in terms of $n_i/(NS)$ and $m_a/(NI)$, respectively. The quantities $NS=\sum_j n_j$ and $NI=\sum_b m_b$ are the total number susceptible and infective individuals respectively, and they are time-dependent themselves. Results are shown in Fig.~\ref{fig:SubgroupCycles}(c) and (d). Although the overall number of infectives, $NI$, undergoes the noise-driven cycles shown in Fig.~\ref{fig:PopulationDynamics}, we find no discernible structure within the group of infectives; the time series $m_a/(NI)$ in Fig.~\ref{fig:SubgroupCycles}(d) are essentially flat noisy lines. This is what one would expect, since the allocation to each subgroup, $I_a$, of infectives is random when an individual is newly infected, and the recovery rate is the same for all infective subgroups.

A more complex behaviour can be seen within the group of susceptibles. This group as a whole undergoes stochastic cycles (see Fig.~\ref{fig:PopulationDynamics}), but an interesting structure is observed within the group of susceptibles as well. The time series $n_i/(NS)$ in Fig.~\ref{fig:SubgroupCycles}(c) show cyclic behaviour, and -- to a good approximation -- any pair of these time series is either in phase with each other, or they have a phase difference of $\pm\pi$. 

To explore the phase lag between the different time series we use the so-called complex coherence function \cite{Stoica2004a}. This technique relies on computing the cross-spectrum $\avg{\widehat x_i(\omega)\widehat x_j^*(\omega)}$ between time series $x_i(t)$ and $x_j(t)$. The phase lag is then obtained as
	\be\label{eq:l}
	{\cal L}_{x_i x_j}(\omega)=\tan^{-1}~\frac{\mbox{Im} \avg{\widehat x_i(\omega)
			\widehat x_j^*(\omega)}}{\mbox{Re} \avg{\widehat x_i(\omega)\widehat x_j^*(\omega)}}.
	\ee
We stress that the subscript ${}^*$ denotes complex conjugation, and is not to be confused with ${}^\star$, used earlier to indicate fixed points of the deterministic dynamics. Eq.~\eqref{eq:l} returns a phase lag for each spectral component, $\omega$. Details can be found in the Supplement (see \ref{app:PhaseLag}). 

\begin{figure*}[t!!]
	\includegraphics[width=0.96\textwidth]{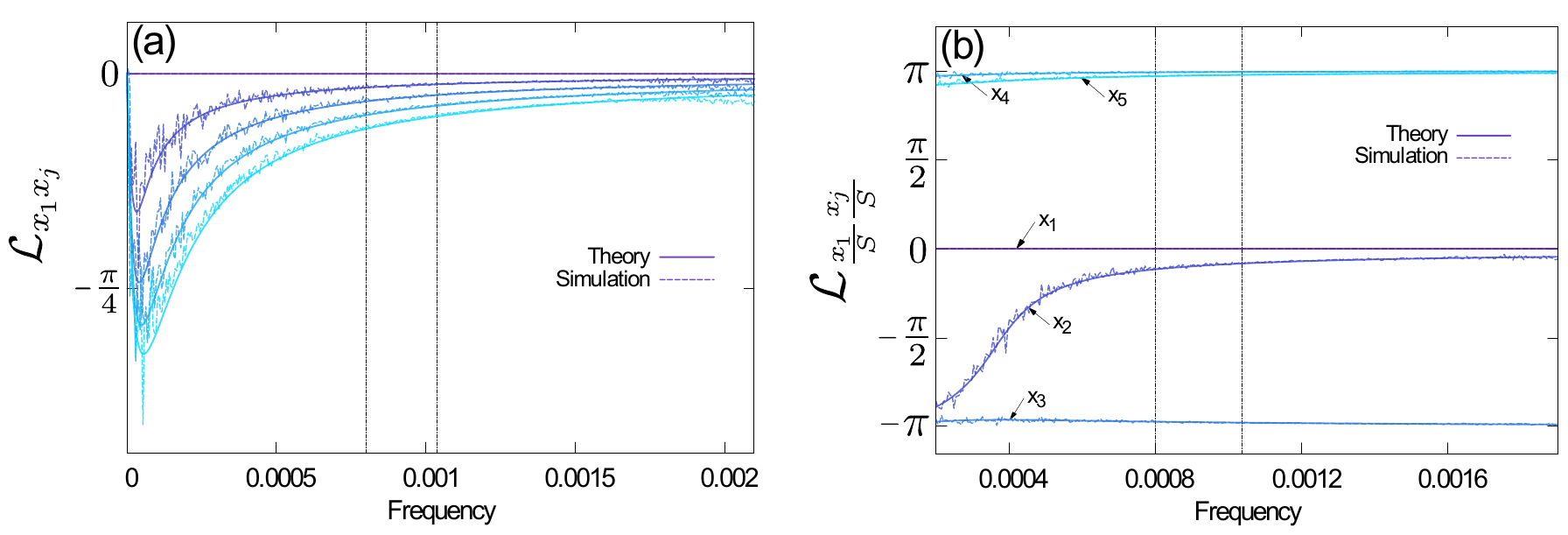}
	\caption{\textbf{Phase-lag of time series between different subgroups of susceptibles.} 
				Data is for the same setup as in Fig.~\ref{fig:SubgroupCycles}. We show the 
				phase-lag between subgroups $i$ and reference subgroup $1$. Panel \textbf{(a)} 
				depicts the case in which time series are normalized with respect to the total 
				population, $N$; in panel \textbf{(b)} input time series are normalized with 
				respect to the total number of susceptibles $NS$. As in 
				Fig.~\ref{fig:PSD_Subclasses}, the vertical dotted lines mark the half-width of 
				the peaks in the corresponding power spectra.}
	\label{fig:PhLg_Si}
	\end{figure*}

The phase lag between the different groups of susceptible individuals is shown in Fig.~\ref{fig:PhLg_Si}. The data in panel (a) corresponds to Fig.~\ref{fig:SubgroupCycles} (a).  More precisely, in Fig.~\ref{fig:PhLg_Si} (a) we pick the time series $n_1/N$ as a reference, and show the phase lag of all subgroups $n_i/N$ with respect to this reference time series. We find that the phase lag for frequencies around the dominant frequency in the power spectra is small, consistent with Fig.~\ref{fig:SubgroupCycles} (a); all time series $n_i/N$ oscillate (roughly) in phase with each other. In Fig.~\ref{fig:PhLg_Si} (b) we repeat this procedure, but now taking the time series $n_i/(NS)$ as an input, corresponding to Fig.~\ref{fig:SubgroupCycles} (c). One then finds a rather different picture; the phase lag around the dominant frequency takes values either near zero, or close to $\pm\pi$.  This indicates that the different classes of susceptible individuals fall into two groups. The time series in either group are in phase with each other, and in anti-phase with those in the respective other group. A closer inspection shows that these two groups are formed by the time series $i$ with $x_{i}^{\star}<S^{\star}/K$ and with $x_i^\star>S^\star/K$ respectively. This behaviour in turn can be understood intuitively by revisiting Eqs.~\eqref{eq:FixedPoint_Si,Ia}. Assuming $\kappa\ll (\xi+\fbeta^\star)\chi_i$ for all $i$ (a valid approximation for the cases analysed here), we find $x_i^\star \propto 1/\chi_i$, indicating that the more susceptible classes are less populated at the deterministic fixed point than the less susceptible ones. During the increasing leg of a stochastic cycle, we expect the number of newly infected individuals among class $i$ to be proportional to $x_i^\star\chi_i$, suggesting that all susceptible classes are depleted in equal absolute numbers. This in turn means that subclasses with $x_i^\star>S^\star/K$ will represent an even larger fraction of the susceptible population as the total susceptible population decreases, while the subclasses with $x_i^\star<S^\star/K$ will represent a smaller fraction. This is what is observed in Fig.~\ref{fig:SubgroupCycles} (c).

		\section{Conclusions}\label{sec:Conclusions}

In summary, we have explored the SIR model in finite populations, including demographic processes and allowed for agent-to-agent heterogeneity in both the susceptibility to a disease and the capacity to spread the disease. This system combines the effects of intrinsic demographic stochasticity (due to random infection, recovery and birth-death events), with quenched heterogeneity. The focus of our paper is to characterise the interplay between these two types of stochasticity, and to investigate how the heterogeneity between individuals affects quasi-cycles driven by intrinsic noise. Our analysis relies on the system-size expansion, which allows us to compute the properties of these cycles analytically in the linear-noise approximation.

Our principal results can be summarised as follows: 
(i)  In the deterministic limit of infinite populations no closed set of equations for macroscopic quantities can be found in the transient regime. Fixed points for aggregate quantities of this deterministic dynamics can however be fully determined from a set of closed equations for the total susceptible ($S^\star$) and infected ($I^\star$) population, and weighted averages of the susceptibility ($\fchi^\star$) and infectivity ($\fbeta^\star$). 
(ii) Similarly, the Langevin equations in the linear-noise approximation do not close easily at the aggregate level, but a closed set of equations for the spectra of fluctuations in $S, I, \fchi$ and $\fbeta$ about the deterministic fixed point can be found in the stationary state. These can be used to analytically describe the stochastic oscillations about the fixed point. 
(iii) Within reasonable assumptions, the characteristic frequency of the noise-driven oscillations is determined mostly by the mean susceptibility and infectivity at birth or infection ($\overline{\chi}$ and $\overline{\beta}$). However, the amplitude of the oscillations and the sharpness of peaks in the power spectra will generally depend on the higher moments of the distribution of susceptibilities and infectivities,in particular also on the agent-to-agent heterogeneity.  
(iv) Finally, the number of individuals in the different subclasses of infectives and susceptibles undergo stochastic cycles as well. If expressed in relation to the total population, these time series are synchronised and in phase. Normalized against the time-dependent total number of infectives, however, the different infective subclasses show no discernible oscillatory behaviour. Using a similar normalization within the susceptible population, we find that different subclasses are syncronized and either in phase with each other or have a phase difference of $\pm \pi$. These results are confirmed analytically. Regardless of the normalization, we find that the periodic outbreaks do not follow a hierarchical infection process, and all subgroups have similar absolute depletion/increase in absolute numbers. This is in contrast to what has been reported in single outbreak studies  \cite{Barthelemy2004, Barthelemy2005}. However, it is important to note that in this existing work the outbreak is tracked in an initial transient period. Our results are valid after this period, at a deterministic fixed point, where the susceptible population is distributed in inverse proportion to their susceptibility (as explained above); this is a scenario different to the one studied in \cite{Barthelemy2004, Barthelemy2005}.

We think our results can be relevant for future work in several ways. 
First, our work contributes to the ongoing discussion about when and how a model with heterogeneity can be replaced or approximated by a homogeneous model. In previous studies, heterogeneous models were compared to homogeneous models with susceptibility equivalent to the arithmetic \cite{Ball1985} or harmonic mean \cite{Andersson1998} of the susceptibilities in the different groups. More recently, the focus has been placed on equivalent basic reproduction numbers ($R_0$) \cite{Yates2006}. In the heterogeneous model this requires estimating $R_0$ based on, for example, the outbreak size, and therefore the comparison is not straightforward. Here we have shown that all models within the class we have looked at and with equal values of $\overline{\chi}\overline{\beta}$ generate periodic outbreaks with the same dominating frequency. This characteristic frequency can be used to define a unique homogeneous model to which models of varying degrees of heterogeneity can be compared.
Furthermore, the dependence of the spectra of oscillations on both the first and higher moments of the distribution of heterogeneity might provide an avenue towards estimating how heterogeneous a population is from the observation of epidemic cycles. Finally, the formalism we have developed is versatile and can be applied to study quasi-cycles in other areas in which heterogeneity might be relevant, for example in predator-prey dynamics or evolution \cite{McKane2005a, Butler2009, Black2012, Bladon2010, Cremer2008, Mobilia2010}. Our findings indicate that the frequency of quasi-cycles can, to a good approximation, be obtained from the first moment of the distribution of heterogeneous agent properties, but that their amplitude depends on higher moments of the disorder. We expect similar behaviour in other heterogeneous systems with noise-driven cycles.


\section*{Acknowledgements}

FHA thanks Consejo Nacional de Ciencia y Tecnolog\'ia (CONACyT, Mexico) for support. TG acknowledges funding by the Engineering and Physical Sciences Research Council (EPSRC, UK) under grant number EP/K037145/1. 	TG would like to thank the Group of Nonlinear Physics, University of Santiago de Compostela, Spain for hospitality.

\section*{Author contributions statement}

FHA and TG conceived and designed the study. FHA carried out the analytical calculations and computer simulations. FHA and TG interpreted results. FHA and TG wrote and reviewed the manuscript.

\section*{Additional information}

The authors have no competing financial interests.

\label{LastPageDoc}		


\clearpage
\setcounter{section}{0}		
\setcounter{page}{1}		
\setcounter{equation}{0}	
\renewcommand{\thesection}{S\arabic{section}} 		
\renewcommand{\thepage}{S\arabic{page}} 			
\renewcommand{\theequation}{S\arabic{equation}}  	
\rfoot{\small\sffamily\bfseries\thepage/\pageref*{LastPageApp}}	

\begin{center}
{\LARGE \bf The effects of heterogeneity \\on stochastic cycles in epidemics\\~\\
Supplemental Material }\\
\vspace{2em}
{\large Francisco Herrer\'{i}as-Azcu\'{e} and Tobias Galla}\\

{\tt francisco.herreriasazcue@postgrad.manchester.ac.uk, tobias.galla@manchester.ac.uk}\\
\vspace{1em}

Theoretical Physics, School of Physics and Astronomy, \\ The University of Manchester, Manchester M13 9PL, United Kingdom
\end{center}


\section{Linear-noise approximation}\label{app:Noise-correlators}

Carrying out the system-size expansion for the model with heterogeneity is tedious, but straightforward and follows the lines of \cite{vanKampen1992}. The final outcome is the linear-noise approximation in Eqs.~(\ref{eq:LNA_Si,Ia}). The variables $\eta_i$ and $\nu_a$, represent Gaussian noise, with no correlation in time, but with potential correlation between the different noise variables at equal time. These noise variables can be decomposed as 
	\BE
	\eta_{i} & = & -\sum_{a}u_{ia}-\sum_{ab}v_{iab}-\sum_{k\neq i}x_{ik}+
					\sum_{k\neq i}x_{ki}+\sum_{a}y_{ai}+z_{i},\nn \\
	\nu_{a} & = & \sum_{i}u_{ia}+\sum_{ib}v_{iba}-w_{a}-\sum_{i}y_{ai},
	\label{eq:noises}
	\EE
	where, broadly speaking, each term on the right-hand side represents one possible type of event in the microscopic model. For example, $u_{ia}$ relates to spontaneous infection of a susceptible individual of type $S_i$, resulting in a newly infective of type $I_a$. Similarly, $v_{iab}$ represents an event in which an individual of type $S_i$ is infected by an individual of type $I_a$, and the newly infected is of type $I_b$. The variable $w_a$ relates to a recovery event of an individual of type $I_a$, death of susceptible $S_i$ and simultaneous birth of susceptible $S_k$ is reflected by $x_{ik}$; death of an individual of type $I_a$ and simultaenous birth of susceptible $S_i$ is described by $y_{ai}$, and finally  death of a recovered individual and simultaneous birth of susceptible $S_i$, by $z_i$. The signs on the right-hand-side in Eqs.~(\ref{eq:noises}) reflect the fact that each of these events may either increase or reduce the number of individuals of type $S_i$ and $I_a$, respectively.
	
Each of the noise variables on the right-hand-side of Eqs.~\eqref{eq:noises} are uncorrelated in time, and they have no cross-correlations. Within the LNA their variances are set by the corresponding reaction rates at the deterministic fixed point, i.e. we have
	\BE
	\bra u_{ia}(t)u_{ia}(t')\ket  	& = & \xi\chi_{i}q_{a}x_{i}^{\star}\delta(t-t'),	\nn \\
	\bra v_{iab}(t)v_{iab}(t')\ket  & = & \beta_{a}\chi_{i}q_{b}x_{i}^{\star}
											I_{a}^{\star}\delta(t-t'),					\nn \\
	\bra w_{a}(t)w_{a}(t')\ket  	& = & \rho I_{a}^{\star}\delta(t-t'),				\nn \\
	\bra x_{ik}(t)x_{ik}(t')\ket  	& = & p_{k}\kappa x_{i}^{\star}\delta(t-t'),			\nn\\
	\bra y_{ai}(t)y_{ai}(t')\ket  	& = & p_{i}\kappa I_{a}^{\star}\delta(t-t'),		\nn \\
	\bra z_{i}(t)z_{i}(t')\ket  	& = & \left(1-S^{\star}-I^{\star}\right)p_i \kappa\delta(t-t') .
	\label{eq:noise_corr_u-z}
	\EE
Using the shorthand introduced in Eqs.~\eqref{eq:AiCDE}, we then find
	\BE
	\label{eq:noisecorr}
	\bra \eta_{i}(t)\eta_{j}(t')\ket  & = & -\kappa^{2}\left(\frac{1}{A_{i}}+
							\frac{1}{A_{j}}\right)p_{i}p_{j}\delta(t-t'), ~\mbox{for $i\neq j$}, \nn \\
	\bra \eta_{i}(t)\eta_{i}(t')\ket  & = & 2\kappa\left(1-\frac{\kappa p_{i}}{A_{i}}\right)
							p_{i}\delta(t-t'),													 \nn \\
	\bra \nu_{a}(t)\nu_{b}(t')\ket  & = & 0,	~\mbox{for $a\neq b$}							 \nn \\
	\bra \nu_{a}(t)\nu_{a}(t')\ket  & = & 2C\fchi^{\star}q_{a}\delta(t-t'),					\nn		 \\
	\bra \eta_{i}(t)\nu_{a}(t')\ket  & = & -\kappa C\left(\frac{\chi_{i}}{A_{i}}+
										\frac{\fchi^{\star}}{D}\right)p_{i}q_{a}\delta(t-t'),	 
	\EE
which are needed for the computation of the PSDs.

\pagebreak

			\section{Calculation of power spectra}\label{app:PSDs}

We start from the result in Eqs.~\eqref{eq:Fourier-AiCDE} in Sect. \ref{sub:Fluctuations}: 
	\BE
	\widehat{S}(\omega) & = & \frac{1}{i\omega+\kappa}\left[-
						\frac{i\omega+D}{\overline{\beta}}\widehat{\fbeta}+\frac{1}
							{\overline{\beta}}\sum_{a}\beta_{a}\widehat{\nu}_a+\sum_{i}
								\widehat{\eta}_i\right],										\nn \\
	\widehat{I}(\omega) & = & \frac{1}{i\omega+D}\left[
						\frac{i\omega+D}{\overline{\beta}}\widehat{\fbeta}-
							\frac{1}{\overline{\beta}}\sum_{a}\beta_{a}\widehat{\nu}_a+
								\sum_{a}\widehat{\nu}_a\right],									\nn \\
	\widehat{\fchi}(\omega)  & = & \frac{1}{\overline{\beta}C}\left[\left(i\omega+E\right)
							\widehat{\fbeta}-\sum_{a}\beta_{a}\widehat{\nu}_a\right],			\nn \\
	\widehat{\fbeta}(\omega) & = & \frac{\overline{\beta}C\sum\limits _{i}\frac{\chi_{i}
								\widehat{\eta}_i}{i\omega+A_{i}}+
								\sum\limits _{a}\beta_{a}\widehat{\nu}_a}{
								i\omega+E+\overline{\beta}C\kappa\sum\limits _{i}
								\frac{\chi_{i}^{2}p_{i}}{A_{i}\left(i\omega+A_{i}\right)}}.
	\EE	
As an illustration let us now compute the power spectrum of $\widehat{\fbeta}$. To keep equations manageable, we define
	\BE
	f_{i}(\omega)&=& \overline{\beta}C\frac{\chi_{i}}{\left(i\omega+A_{i}\right)}, \nn \\	
	g(\omega)&=&\left(i\omega+E\right)+\overline{\beta}C\kappa\sum\limits _{i}
		\frac{\chi_{i}^{2}p_{i}}{A_{i}\left(i\omega+A_{i}\right)},
 	\EE
and so we write the Fourier transform of $\tilde{\fbeta}$ as
	\BE
	 \widehat{\fbeta}(\omega)&=&\frac{\sum\limits _{i}f_{i}\widehat{\eta}_i+
		\sum\limits _{a}\beta_{a}\widehat{\nu}_a}{g},
	\label{eq:psihat(f,g)}
	\EE
where $f_i$, $\beta_a$, $\widehat{\eta}_i$ and $\widehat{\nu}_a$ are all functions of $\omega$. We then find 
	\BE
	\mathcal{P}_{\fbeta}(\omega)  & = & \bra \left(\frac{\sum_{i}f_{i}\widehat{\eta}_i+
							\sum_{a}\beta_{a}\widehat{\nu}_a}{g}\right)
							\left(\frac{\sum_{i}f_{i}^*\widehat{\eta}_i+
							\sum_{a}\beta_{a}\widehat{\nu}_a}{g^*} \right)
							\ket  \nn \\
						& = & \frac{1}{|g|^{2}}\left(\sum\limits _{i,j}
							f_{i}f_{j}^{*}\bra \widehat{\eta}_i
							\widehat{\eta}_{j}\ket +\sum\limits _{i,b}f_{i}
							\beta_{b}\bra \widehat{\eta}_i\widehat{\nu_{b}}\ket +
							\sum\limits _{a,j}f_{j}^{*}\beta_{a}\bra 
							\widehat{\eta}_{j}\widehat{\nu}_a\ket +\sum\limits _{a,b}
							\beta_{a}\beta_{b}\bra \widehat{\nu}_a
							\widehat{\nu_{b}}\ket \right)						\nn \\
	 					& = & \frac{1}{|g|^{2}}\left(\sum\limits _{i}
	 						f_{i}f_{i}^{*}\bra \widehat{\eta}_i\widehat{\eta}_i\ket 
	 						+\sum\limits _{i}\sum\limits _{j\neq i}f_{i}f_{j}^{*}\bra
	 						 \widehat{\eta}_i\widehat{\eta}_{j}\ket +
	 						 \sum\limits _{i,b}\left(f_{i}+f_{i}^{*}\right)
	 						 \beta_{b}\bra \widehat{\eta}_i\widehat{\nu_{b}}\ket +
	 						 \sum\limits _{a}\beta_{a}^{2}\bra \widehat{\nu}_a
	 						 \widehat{\nu}_a\ket \right).
	\EE
The notation $*$ denotes complex conjugation. Substituting the noise correlators from Eqs.~\eqref{eq:noisecorr}, 
	\BE
	\mathcal{P}_{\fbeta}(\omega)	& = & \frac{1}{|g|^{2}}\left(2\kappa\sum\limits _{i}
									f_{i}f_{i}^{*}p_{i}-\kappa^{2}\sum\limits _{i,j}f_{i}f_{j}^{*}
									\left(\frac{1}{A_{i}}+\frac{1}{A_{j}}\right)p_{i}p_{j}				
									\right. \nn \\ & & \left. \hspace{10em}		
	 								-\overline{\beta}\kappa C\sum\limits _{i}\left(
	 								f_{i}+f_{i}^{*}\right)\left(\frac{\chi_{i}}{A_{i}}+
	 								\frac{\phi^{\star}}{D}\right)p_{i}+2
	 								\overline{\beta^{2}}C\phi^{\star}\right),
	\EE
and, using Eq.~\eqref{eq:psihat(f,g)}, we find
	\BE
	\mathcal{P}_{\fbeta}(\omega)	& = & \frac{2\phi^{\star}C}{|g|^{2}}\left(\overline{\beta^{2}}-
							\frac{\overline{\beta}^{2}C\kappa}{D}
							\sum\limits _{i}\frac{\chi_{i}p_{i}A_{i}}{
							\omega^{2}+A_{i}^{2}}\right)
	 						-\frac{\left(\overline{\beta}C\kappa\right)^{2}}{|g|^{2}}
	 						\sum\limits _{i,j}\frac{\chi_{i}p_{i}\chi_{j}p_{j}\left(A_{i}+A_{j}\right)
	 						\left(\omega^{2}+A_{i}A_{j}\right)}
	 						{A_{i}A_{j}(\omega^{2}+A_{i}^{2})(\omega^{2}+A_{j}^{2})},
	\EE
which is the PSD of $\fbeta$, as also reported in Eq.~\eqref{eq:PSD_psi} in the main text. 

Following the same process, we can compute the PSD for the remaining quantities, $\fchi$, $I$ and $S$. We do not report all details, but only the final results 
	\BE\label{eq:PSD}
	\mathcal{P}_{\fchi}(\omega) & = & \frac{1}{\overline{\beta}^{2}C}
						\left[2\overline{\beta^{2}}\fchi^{\star}+\left(\omega^{2}+E^{2}\right)
						\left(\frac{\mathcal{P}_{\fbeta}}{C}-\frac{4\overline{\beta^{2}}
						\fchi^{\star}}{|g|^{2}}\right)\right]								\nn	\\
	 					&   & +\frac{2\kappa}{\overline{\beta}|g|^{2}}\sum\limits _{i}
	 					\frac{\chi_{i}^{2}p_{i}}
	 					{A_{i}\left(\omega^{2}+A_{i}^{2}\right)}
	 					\left[2\fchi^{\star}\overline{\beta^{2}}\left(\omega^{2}-EA_{i}\right)+
	 					\overline{\beta}A_{i}\left(\frac{A_{i}\fchi^{\star}}{D\chi_{i}}+1\right)
	 					\left(\omega^{2}+E^{2}\right)\right]								\nn	\\
	 					&   & +\frac{2\overline{\beta}C\kappa^{2}}{|g|^{2}}
	 					\sum\limits _{i,j}\frac{\chi_{i}^{2}p_{i}\chi_{j}^{2}p_{j}\left[E
	 					\left(\omega^{2}+A_{i}A_{j}\right)+\omega^{2}\left(A_{j}-A_{i}\right)\right]}
	 					{A_{i}A_{j}\left(\omega^2+A_{i}^2\right)\left(\omega^2+A_{j}^2\right)}
	 					\left(\frac{A_{i}\fchi^{\star}}{D\chi_{i}}+1\right),					\nn	\\
	\nn\\\nn\\
	\mathcal{P}_{I}(\omega) & = & \frac{\mathcal{P}_{\fbeta}}{\overline{\beta}^{2}}+
						\frac{4C\fchi^{\star}\left(\overline{\beta}^{2}-\overline{\beta^{2}}
						\right)\left(DE+\omega^{2}\right)}
						{|g|^{2}\overline{\beta}^{2}\left(\omega^{2}+D^{2}\right)}
	 					+\frac{4\kappa C^{2}\fchi^{\star}\left(\overline{\beta}^{2}-
	 					\overline{\beta^{2}}\right)}{\left(|g|^{2}\right)\left(
	 					\omega^{2}+D^{2}\right)\left(\overline{\beta}\right)}
	 					\sum\limits _{i}\frac{\chi_{i}^{2}p_{i}\left(DA_{i}-\omega^{2}\right)}
	 					{A_{i}\left(\omega^{2}+A_{i}^{2}\right)},							\nn	\\
	\nn\\\nn\\
	\mathcal{P}_{S}(\omega) & = & \frac{1}{\omega^{2}+\kappa^{2}}
						\left[\left(\frac{\omega^{2}+D^{2}}{\overline{\beta}^{2}}\right)
						\mathcal{P}_{\fbeta}+2\kappa\left(1-\kappa\sum\limits _{i}
						\frac{p_{i}}{A_{i}}\right)-C\kappa\left(\frac{\fchi^{\star}}{D}+
						\sum\limits _{i}\frac{\chi_{i}p_{i}}{A_{i}}\right)\right]			\nn \\
	 					&  & +\frac{2C}{|g|^{2}\left(\omega^{2}+\kappa^{2}\right)}
	 					\left[-\frac{2\fchi^{\star}\overline{\beta^{2}}}{\overline{\beta}^{2}}+\kappa
	 					\left(\frac{\fchi^{\star}}{D}+\sum_{i}\frac{\chi_{i}p_{i}}{A_{i}}\right)\right]
	 					\left[DE+\omega^{2}-\frac{|g|^{2}}{2}+\overline{\beta}C\kappa
	 					\sum\limits _{j}\frac{\chi_{j}^{2}p_{j}\left(DA_{j}-\omega^{2}\right)}
	 					{A_{j}\left(\omega^{2}+A_{j}^{2}\right)}\right]						\nn \\
	 					&  & +\frac{2C\kappa}{|g|^{2}\left(\omega^{2}+\kappa^{2}\right)}
	 					\sum\limits _{i}\frac{\chi_{i}p_{i}}{A_{i}\left(\omega^{2}+A_{i}^{2}\right)}
	 					\left[C\left(\frac{A_{i}\fchi^{\star}}{D}+\chi_{i}\right)-2A_{i}+\kappa
	 					\left(1+A_{i}\sum_{k}\frac{p_{k}}{A_{k}}\right)\right]				\nn \\
	 					&  & \hspace{6em}\left\{A_{i}DE+\omega^{2}\left(A_{i}+E-D\right)+
	 					\overline{\beta}C\kappa\sum\limits _{j}\frac{\chi_{j}^{2}p_{j}
	 					\left[D\left(A_{i}A_{j}+\omega^{2}\right)+\omega^{2}\left(A_{j}-A_{i}
	 					\right)\right]}{A_{j}\left(\omega^{2}+A_{j}^{2}\right)}\right\}.		
	\EE
The power spectra of fluctuations for the individual subgroups of infectives and susceptibles are found as 
	\BE
	\mathcal{P}_{x_{i}}(\omega) & = & \frac{1}{\omega^{2}+A_{i}^{2}}
						\left[\left(\frac{\kappa\chi_{i}p_{i}}{A_{i}}\right)^{2}\mathcal{P}_{\fbeta}
						+2\kappa\left(1-\frac{\kappa p_{i}}{A_{i}}\right)p_{i}\right]		\nn \\
	 					&  & +\frac{2\overline{\beta}C\kappa^{2}\chi_{i}p_{i}^2}
	 					{|g|^{2}A_{i}\left(\omega^{2}+A_{i}^{2}\right)}
	 					\left[\frac{\fchi^{\star}}{D}+\frac{\chi_{i}\left(\omega^{2}-A_{i}^{2}\right)}
	 					{A_{i}\left(\omega^{2}+A_{i}^{2}\right)}+\frac{\kappa}{A_{i}}
	 					\sum\limits _{j}\frac{\chi_{j}p_{j}\left(A_{i}+A_{j}\right)}
	 					{\omega^{2}+A_{j}^{2}}\right]\left(E+\sum\limits _{k}\frac{\overline{\beta}
	 					C\kappa\chi_{k}^{2}p_{k}}{\omega^{2}+A_{k}^{2}}\right)				\nn \\
	 					&  & +\frac{2\overline{\beta}C\kappa^{2}\chi_{i}p_{i}^2\omega^{2}}
	 					{|g|^{2}A_{i}\left(\omega^{2}+A_{i}^{2}\right)}
	 					\left[\frac{2\chi_{i}}{\omega^{2}+A_{i}^{2}}-\frac{\kappa}{A_{i}}
	 					\sum\limits _{j}\frac{\chi_{j}p_{j}\left(A_{i}+A_{j}\right)}{A_{j}
	 					\left(\omega^{2}+A_{j}^{2}\right)}\right]
	 					\left(1-\sum\limits _{k}\frac{\overline{\beta}C\kappa\chi_{k}^{2}p_{k}}
	 					{A_{k}\left(\omega^{2}+A_{k}^{2}\right)}\right),					\\
	\nn \\ \nn \\
	\mathcal{P}_{y_{a}}(\omega) & = & q_{a}^{2}\mathcal{P}_{I}+\frac{2C\fchi^{\star}q_{a}
						\left(1-q_{a}\right)}{\omega^{2}+D^{2}}.
	\EE

\pagebreak
		\section{Phase Lag}\label{app:PhaseLag}

In order to explore the the phase lag we use the so-called complex coherence function, $\mathcal{CCF}_{ij}$, between subgroups $i$ and $j$, defined as 
	\be
	\mathcal{CCF}_{ij}(\omega)=\frac{\bra \widehat{x}_i\widehat{x}_{j}^{*}\ket }
					{\sqrt{\bra \widehat{x}_i\widehat{x}_i^{*}\ket 
					\bra \widehat{x}_{j}\widehat{x}_{j}^{*}\ket }}=
					\frac{\mathcal{P}_{x_{i}x_{j}}}{\sqrt{\mathcal{P}_{x_{i}}\mathcal{P}_{x_{j}}}},
	\ee
where $\widehat{x}_i$ and $\mathcal{P}$ are functions of $\omega$.

For $i\neq j$ this is in general a complex-valued function (of $\omega$). The argument of $\mathcal{CCF}_{ij}$, given by
	\be
	\mathfrak{L}_{x_i x_j}(\omega)=\tan^{-1}~\frac{\mbox{Im}~ \mathcal{CCF}_{ij}(\omega)}
				{\mbox{Re}~\mathcal{CCF}_{ij}(\omega)}=\tan^{-1}~\frac{\mbox{Im}~
				\mathcal{P}_{x_{i}x_{j}}(\omega)}{\mbox{Re}~\mathcal{P}_{x_{i}x_{j}}(\omega)},
	\label{eq:PhaseLag-2}
	\ee
is known as the phase spectrum; it describes the phase-lag between the time series $x_{i}(t)$ and $x_{j}(t)$ \cite{Rozhnova2012}.

The cross spectra of the population in the susceptible classes normalized with respect to the total population ($x_i=n_{i}/N$) is given by 
	\be
	\mathcal{P}_{x_{i}x_{j}}(\omega)=\bra \widehat{x}_i\widehat{x}_{j}^{*}\ket  = 
	 		\bra \left(\frac{-\chi_{i}x_{i}^{\star}\widehat{\fbeta}+
	 		\widehat{\eta}_i}{i\omega+A_{i}}\right)\left(\frac{-\chi_{j}
	 		x_{j}^{\star}\widehat{\fbeta}^{*}+\widehat{\eta}_{j}}{-i\omega+A_{j}
	 		}\right)\ket.
	\ee
This can be written as
	\be
	\mathcal{P}_{x_{i}x_{j}}(\omega) = \frac{\left(\omega^{2}+A_{i}A_{j}\right)\mathcal{W}_{ij}-
							\omega\left(A_{i}-A_{j}\right)\mathcal{U}_{ij}}
							{\left(\omega^{2}+A_{i}^{2}\right)\left(\omega^{2}+A_{j}^{2}\right)}
	 						+i ~ \frac{\left(\omega^{2}+A_{i}A_{j}\right)
	 						\mathcal{U}_{ij}+\omega\left(A_{i}-A_{j}\right)\mathcal{W}_{ij}}
	 						{\left(\omega^{2}+A_{i}^{2}\right)\left(\omega^{2}+A_{j}^{2}\right)},
	\ee
where we introduced the notation
	\BE
	\mathcal{U}_{ij}(\omega) & = & \chi_{j}x_{j}^{\star}\mbox{Im}\bra 
				\widehat{\eta}_i\widehat{\fbeta}\ket -\chi_{i}x_{i}^{\star}\mbox{Im}
				\bra \widehat{\eta}_{j}\widehat{\fbeta}\ket , 								\nn \\
	\mathcal{W}_{ij}(\omega) & = & \left(\chi_{i}\chi_{j}x_{i}^{\star}x_{j}^{\star}\right)
				\mathcal{P}_{\fbeta}+\bra \widehat{\eta}_{i}\widehat{\eta}_{j}\ket 
				-\chi_{j}x_{j}^{\star}\mbox{Re}\bra \widehat{\eta}_i\widehat{\fbeta}\ket 
				-\chi_{i}x_{i}^{\star}\mbox{Re}\bra \widehat{\eta}_{j}\widehat{\fbeta}\ket.
	\EE
From these we obtain the phase lag as
	\be
	\mathfrak{L}_{x_i x_j}(\omega)=\tan^{-1}~\frac{\omega\left(
				A_{i}-A_{j}\right)\mathcal{W}_{ij}+\left(\omega^{2}+A_{i}A_{j}\right)
				\mathcal{U}_{ij}}{\left(\omega^{2}+
				A_{i}A_{j}\right)\mathcal{W}_{ij}-\omega\left(A_{i}-A_{j}\right)
				\mathcal{U}_{ij}},
	\label{eq:PhaseLag-Sij}
	\ee
which yields the theoretical lines in Fig.~\ref{fig:PhLg_Si}a.

To explore the phase lag between the susceptible subgroups when normalized by the total susceptible population ($x_{i}'=n_{i}/NS$), we first need to compute the cross-spectra of the renormalized signals $\mathcal{P}_{x_{i}'x_{j}'}(\omega)$. As in Section \ref{sub:stoch}, we start from the ansatz
	\be
	\frac{n_i}{NS}=x_{i}'+\frac{1}{\sqrt{N}}\tilde x_{i}'.
	\ee
We then have  
	\be
	\frac{n_i}{NS} = \frac{n_i/N}{S} 
				   = \frac{x_{i}+\frac{1}{\sqrt{N}}\tilde{x}_{i}}{S+\frac{1}{\sqrt{N}}\tilde S}
				   \equiv x_{i}'+\frac{1}{\sqrt{N}}\tilde x_{i}',
				   					   \ee
and so (after expanding in $1/\sqrt{N}$)
				   
\be
				   \tilde x'_{i}=\frac{S^{\star}\,\tilde{x_{i}}-x_{i}^{\star}\,\tilde{S}}
				   {\left(S^{\star}\right)^2}.
	\ee
In Fourier space this turns into
	\be
	\widehat{x}_i'=\frac{S^{\star}\,\widehat{x}_i-x_{i}^{\star}\,\widehat{S}}
				{\left(S^{\star}\right)^{2}}.
	\ee
For the cross spectra we then find
	\be
	\mathcal{P}_{x_{i}'x_{j}'}(\omega)=\bra \widehat{x}_i'\widehat{x}_{j}'^{*}\ket   = 
		\bra \frac{\left(S^{\star}\widehat{x}_i-x_{i}^{\star}\widehat{S}\right)
		\left(S^{\star}\widehat{x}_{j}^{*}-x_{j}^{\star}\,\widehat{S}^{*}\right)}
		{\left(S^{\star}\right)^{4}}\ket ,
	\ee
which can be rewritten as
	\be
	\mathcal{P}_{x_{i}'x_{j}'}(\omega)=\frac{1}{\left(S^{\star}\right)^{3}}
					\left(S^{\star}\mbox{Re}\left[\mathcal{P}_{x_{i}x_{j}}\right]
					+\frac{\kappa^{2}p_{i}p_{j}}{S^{\star}A_{i}A_{j}}\mathcal{P}_{S}-\frac{Y_{ij\,R}
					+Y_{ji\,R}}{\overline{\beta}\left(\omega^{2}+\kappa^{2}\right)}\right)
					+\frac{i}{\left(S^{\star}\right)^{3}}\left(S^{\star}\mbox{Im}\left[
					\mathcal{P}_{x_i x_j}\right]-\frac{Y_{ij\,I}-Y_{ji\,I}}{\overline{\beta}
					\left(\omega^{2}+\kappa^{2}\right)}\right).
	\ee
We have  introduced the notation $Y_{ij\,R}=\mbox{Re}\left[Y_{ij}\right]$ and $Y_{ij\,I}=\mbox{Im}\left[Y_{ij}\right]$ with
	\BE
	Y_{ij}(\omega) & = & \frac{\left(\omega^{2}+i\omega\left(\kappa-A_{j}\right)+
					\kappa A_{j}\right)\kappa p_{i}}{\left(\omega^{2}+A_{j}^{2}\right)A_{i}}
					\left\{\frac{\kappa\chi_{j}p_{j}}{A_{j}}\left(D\mathcal{P}_{\fbeta}-\sum_{a}
					\beta_{a}\mbox{Re}\bra \widehat{\nu}_a\widehat{\fbeta}\ket -\overline{\beta}
					\sum\limits _{k}\mbox{Re}\bra \widehat{\eta_{k}}\widehat{\fbeta}\ket 
					\right)\qquad\right.												\nn \\
	 		  &   & \hspace{4em}\left.\omega \mbox{Im}\bra\widehat{\eta}_j\widehat{\fbeta}\ket 
	 				-D\mbox{Re}\bra \widehat{\eta}_{j}\widehat{\fbeta}\ket 
	 				-\kappa\overline{\beta}p_{j}\left[C\left(\frac{\chi_{j}}{A_{j}}+
	 				\frac{\fchi^{\star}}{D}\right)-2+\frac{\kappa}{A_{j}}+
	 				\kappa\sum\limits _{k}\left(\frac{p_{k}}{A_{k}}\right)\right]\right\}	\nn \\
	 		  &   & +i\frac{\left[\omega^{2}+i\omega\left(\kappa-A_{j}\right)+\kappa A_{j}\right]
	 		  		\kappa p_{i}}{\left(\omega^{2}+A_{j}^{2}\right)A_{i}}
	 		  		\left[\frac{\kappa\chi_{j}p_{j}}{A_{j}}\left(\omega\mathcal{P}_{\fbeta}+\sum_{a}
	 		  		\beta_{a}\mbox{Im}\bra \widehat{\nu}_a\widehat{\fbeta}\ket +\overline{\beta}
	 		  		\sum\limits _{k}\mbox{Im}\bra \widehat{\eta_{k}}\widehat{\fbeta}\ket 
	 		  		\right)\qquad\right.												\nn \\
	 		  &   & \hspace{16em}\left.-\omega \mbox{Re}\bra \widehat{\eta}_{j}\widehat{\fbeta}\ket
	 		  		-D\,\mbox{Im}\bra \widehat{\eta}_{j}\widehat{\fbeta}\ket 
	 		  		\vphantom{\sum\limits _{k}\frac{\chi_j}{A_j}}\right].
	\EE
From these, we can find the phase-lag as
	\be
	\mathfrak{L}_{x'_i x'_j}(\omega)=\tan^{-1}~\frac{S^{\star}
					\mbox{Im}\left[\mathcal{P}_{x_{i}x_{j}}\right]-\frac{Y_{ij\,I}-Y_{ji\,I}}
					{\overline{\beta}\left(\omega^{2}+\kappa^{2}\right)}}{S^{\star}
					\mbox{Re}\left[\mathcal{P}_{x_{i}x_{j}}\right]+\frac{\kappa^{2}p_{i}p_{j}}
					{S^{\star}A_{i}A_{j}}\mathcal{P}_{S}-\frac{Y_{ij\,R}+Y_{ji\,R}}
					{\overline{\beta}\left(\omega^{2}+\kappa^{2}\right)}}.
	\label{eq:PhaseLag-sij}
	\ee

This expression was used to obtain the analytical predictions shown in Fig.~\ref{fig:PhLg_Si}b.

\label{LastPageApp}		



\begin{thebibliography}{10}
\expandafter\ifx\csname url\endcsname\relax
  \def\url#1{\texttt{#1}}\fi
\expandafter\ifx\csname urlprefix\endcsname\relax\def\urlprefix{URL }\fi
\expandafter\ifx\csname doiprefix\endcsname\relax\def\doiprefix{DOI }\fi
\providecommand{\bibinfo}[2]{#2}
\providecommand{\eprint}[2][]{\url{#2}}

\bibitem{Murray2002}
\bibinfo{author}{Murray, J.~D.}
\newblock \emph{\bibinfo{title}{{Mathematical biology}}},
  vol.~\bibinfo{volume}{17} of \emph{\bibinfo{series}{Interdisciplinary Applied
  Mathematics}} (\bibinfo{publisher}{Springer-Verlag}, \bibinfo{address}{Berlin
  Heidelberg}, \bibinfo{year}{2002}), \bibinfo{edition}{3rd} edn.

\bibitem{Wilkinson2011}
\bibinfo{author}{Wilkinson, D.~J.}
\newblock \emph{\bibinfo{title}{{Stochastic Modelling for Systems Biology}}}
  (\bibinfo{publisher}{CRC Press}, \bibinfo{address}{Boca Raton},
  \bibinfo{year}{2011}), \bibinfo{edition}{2nd} edn.

\bibitem{Goel1974}
\bibinfo{author}{Goel, N.~S.} \& \bibinfo{author}{Richter-Dyn, N.}
\newblock \emph{\bibinfo{title}{{Stochastic Models in Biology}}}
  (\bibinfo{publisher}{Academic Press}, \bibinfo{address}{New York, NY},
  \bibinfo{year}{1974}).

\bibitem{Andersson2000}
\bibinfo{author}{Andersson, H.} \& \bibinfo{author}{Britton, T.}
\newblock \emph{\bibinfo{title}{{Stochastic epidemic models and their
  statistical analysis}}}, vol. \bibinfo{volume}{151} of
  \emph{\bibinfo{series}{Lecture Notes in Statistics}}
  (\bibinfo{publisher}{Springer New York}, \bibinfo{address}{New York, NY},
  \bibinfo{year}{2000}).

\bibitem{Elowitz2011}
\bibinfo{author}{Elowitz, M.~B.}
\newblock \bibinfo{journal}{\bibinfo{title}{{Stochastic gene expression in a
  single cell}}}.
\newblock {\emph{\JournalTitle{Science}}} \textbf{\bibinfo{volume}{297}},
  \bibinfo{pages}{1183--1186} (\bibinfo{year}{2002}).
\newblock \doiprefix 10.1126/science.1070919.

\bibitem{Paulsson2004}
\bibinfo{author}{Paulsson, J.}
\newblock \bibinfo{journal}{\bibinfo{title}{{Summing up the noise in gene
  networks.}}}
\newblock {\emph{\JournalTitle{Nature}}} \textbf{\bibinfo{volume}{427}},
  \bibinfo{pages}{415--8} (\bibinfo{year}{2004}).
\newblock \doiprefix 10.1038/nature02257.

\bibitem{Moreno2002}
\bibinfo{author}{Moreno, Y.}, \bibinfo{author}{Pastor-Satorras, R.} \&
  \bibinfo{author}{Vespignani, A.}
\newblock \bibinfo{journal}{\bibinfo{title}{{Epidemic outbreaks in complex
  heterogeneous networks}}}.
\newblock {\emph{\JournalTitle{The European Physical Journal B}}}
  \textbf{\bibinfo{volume}{26}}, \bibinfo{pages}{521--529}
  (\bibinfo{year}{2002}).
\newblock \doiprefix 10.1007/s10051-002-8996-y.

\bibitem{Raj2008}
\bibinfo{author}{Raj, A.} \& \bibinfo{author}{van Oudenaarden, A.}
\newblock \bibinfo{journal}{\bibinfo{title}{{Nature, nurture, or chance:
  Stochastic gene expression and its consequences}}}.
\newblock {\emph{\JournalTitle{Cell}}} \textbf{\bibinfo{volume}{135}},
  \bibinfo{pages}{216--226} (\bibinfo{year}{2008}).
\newblock \doiprefix 10.1016/j.cell.2008.09.050.

\bibitem{Heldt2015}
\bibinfo{author}{Heldt, F.~S.}, \bibinfo{author}{Kupke, S.~Y.},
  \bibinfo{author}{Dorl, S.}, \bibinfo{author}{Reichl, U.} \&
  \bibinfo{author}{Frensing, T.}
\newblock \bibinfo{journal}{\bibinfo{title}{{Single-cell analysis and
  stochastic modelling unveil large cell-to-cell variability in influenza A
  virus infection}}}.
\newblock {\emph{\JournalTitle{Nature Communications}}}
  \textbf{\bibinfo{volume}{6}}, \bibinfo{pages}{8938} (\bibinfo{year}{2015}).
\newblock \doiprefix 10.1038/ncomms9938.

\bibitem{Scott2006}
\bibinfo{author}{Scott, M.}, \bibinfo{author}{Ingalls, B.} \&
  \bibinfo{author}{K{\ae}rn, M.}
\newblock \bibinfo{journal}{\bibinfo{title}{{Estimations of intrinsic and
  extrinsic noise in models of nonlinear genetic networks}}}.
\newblock {\emph{\JournalTitle{Chaos: An Interdisciplinary Journal of Nonlinear
  Science}}} \textbf{\bibinfo{volume}{16}}, \bibinfo{pages}{026107}
  (\bibinfo{year}{2006}).
\newblock \doiprefix 10.1063/1.2211787.

\bibitem{Swain2002}
\bibinfo{author}{Swain, P.~S.}, \bibinfo{author}{Elowitz, M.~B.} \&
  \bibinfo{author}{Siggia, E.~D.}
\newblock \bibinfo{journal}{\bibinfo{title}{{Intrinsic and extrinsic
  contributions to stochasticity in gene expression}}}.
\newblock {\emph{\JournalTitle{Proceedings of the National Academy of
  Sciences}}} \textbf{\bibinfo{volume}{99}}, \bibinfo{pages}{12795--12800}
  (\bibinfo{year}{2002}).
\newblock \doiprefix 10.1073/pnas.162041399.

\bibitem{Hilfinger2011a}
\bibinfo{author}{Hilfinger, A.} \& \bibinfo{author}{Paulsson, J.}
\newblock \bibinfo{journal}{\bibinfo{title}{{Separating intrinsic from
  extrinsic fluctuations in dynamic biological systems}}}.
\newblock {\emph{\JournalTitle{Proceedings of the National Academy of
  Sciences}}} \textbf{\bibinfo{volume}{108}}, \bibinfo{pages}{12167--12172}
  (\bibinfo{year}{2011}).
\newblock \doiprefix 10.1073/pnas.1018832108.

\bibitem{Alonso2007}
\bibinfo{author}{Alonso, D.}, \bibinfo{author}{McKane, A.~J.} \&
  \bibinfo{author}{Pascual, M.}
\newblock \bibinfo{journal}{\bibinfo{title}{{Stochastic amplification in
  epidemics}}}.
\newblock {\emph{\JournalTitle{Journal of The Royal Society Interface}}}
  \textbf{\bibinfo{volume}{4}}, \bibinfo{pages}{575--582}
  (\bibinfo{year}{2007}).
\newblock \doiprefix 10.1098/rsif.2006.0192.

\bibitem{Olsen1990}
\bibinfo{author}{Olsen, L.} \& \bibinfo{author}{Schaffer, W.}
\newblock \bibinfo{journal}{\bibinfo{title}{{Chaos versus noisy periodicity:
  alternative hypotheses for childhood epidemics}}}.
\newblock {\emph{\JournalTitle{Science}}} \textbf{\bibinfo{volume}{249}},
  \bibinfo{pages}{499--504} (\bibinfo{year}{1990}).
\newblock \doiprefix 10.1126/science.2382131.

\bibitem{Black2009}
\bibinfo{author}{Black, A.~J.}, \bibinfo{author}{McKane, A.~J.},
  \bibinfo{author}{Nunes, A.} \& \bibinfo{author}{Parisi, A.}
\newblock \bibinfo{journal}{\bibinfo{title}{{Stochastic fluctuations in the
  susceptible-infective-recovered model with distributed infectious periods}}}.
\newblock {\emph{\JournalTitle{Physical Review E}}}
  \textbf{\bibinfo{volume}{80}}, \bibinfo{pages}{021922}
  (\bibinfo{year}{2009}).
\newblock \doiprefix 10.1103/PhysRevE.80.021922.

\bibitem{Rozhnova2009}
\bibinfo{author}{Rozhnova, G.} \& \bibinfo{author}{Nunes, A.}
\newblock \bibinfo{journal}{\bibinfo{title}{{Fluctuations and oscillations in a
  simple epidemic model}}}.
\newblock {\emph{\JournalTitle{Physical Review E}}}
  \textbf{\bibinfo{volume}{79}}, \bibinfo{pages}{041922}
  (\bibinfo{year}{2009}).
\newblock \doiprefix 10.1103/PhysRevE.79.041922.

\bibitem{McKane2005a}
\bibinfo{author}{McKane, A.~J.} \& \bibinfo{author}{Newman, T.~J.}
\newblock \bibinfo{journal}{\bibinfo{title}{{Predator-prey cycles from resonant
  amplification of demographic stochasticity}}}.
\newblock {\emph{\JournalTitle{Physical Review Letters}}}
  \textbf{\bibinfo{volume}{94}}, \bibinfo{pages}{1--4} (\bibinfo{year}{2005}).
\newblock \doiprefix 10.1103/PhysRevLett.94.218102.

\bibitem{Bjornstad2001}
\bibinfo{author}{Bj{\o}rnstad, O.~N.} \& \bibinfo{author}{Grenfell, B.~T.}
\newblock \bibinfo{journal}{\bibinfo{title}{{Noisy clockwork: time series
  analysis of population fluctuations in animals.}}}
\newblock {\emph{\JournalTitle{Science}}} \textbf{\bibinfo{volume}{293}},
  \bibinfo{pages}{638--643} (\bibinfo{year}{2001}).
\newblock \doiprefix 10.1126/science.1062226.

\bibitem{Bladon2010}
\bibinfo{author}{Bladon, A.~J.}, \bibinfo{author}{Galla, T.} \&
  \bibinfo{author}{McKane, A.~J.}
\newblock \bibinfo{journal}{\bibinfo{title}{{Evolutionary dynamics, intrinsic
  noise, and cycles of cooperation}}}.
\newblock {\emph{\JournalTitle{Physical Review E}}}
  \textbf{\bibinfo{volume}{81}} (\bibinfo{year}{2010}).
\newblock \doiprefix 10.1103/PhysRevE.81.066122.

\bibitem{Samoilov2005}
\bibinfo{author}{Samoilov, M.}, \bibinfo{author}{Plyasunov, S.} \&
  \bibinfo{author}{Arkin, A.~P.}
\newblock \bibinfo{journal}{\bibinfo{title}{{Stochastic amplification and
  signaling in enzymatic futile cycles through noise-induced bistability with
  oscillations}}}.
\newblock {\emph{\JournalTitle{Proceedings of the National Academy of
  Sciences}}} \textbf{\bibinfo{volume}{102}}, \bibinfo{pages}{2310--2315}
  (\bibinfo{year}{2005}).
\newblock \doiprefix 10.1073/pnas.0406841102.

\bibitem{Bolker1993}
\bibinfo{author}{Bolker, B.~M.} \& \bibinfo{author}{Grenfell, B.~T.}
\newblock \bibinfo{journal}{\bibinfo{title}{{Chaos and biological complexity in
  measles dynamics}}}.
\newblock {\emph{\JournalTitle{Proceedings of the Royal Society B: Biological
  Sciences}}} \textbf{\bibinfo{volume}{251}}, \bibinfo{pages}{75--81}
  (\bibinfo{year}{1993}).
\newblock \doiprefix 10.1098/rspb.1993.0011.

\bibitem{Schenzle1984}
\bibinfo{author}{Schenzle, D.}
\newblock \bibinfo{journal}{\bibinfo{title}{{An age-structured model of pre-
  and post-vaccination measles transmission}}}.
\newblock {\emph{\JournalTitle{Mathematical Medicine and Biology}}}
  \textbf{\bibinfo{volume}{1}}, \bibinfo{pages}{169--191}
  (\bibinfo{year}{1984}).
\newblock \doiprefix 10.1093/imammb/1.2.169.

\bibitem{Earn2000}
\bibinfo{author}{Earn, D. J.~D.}, \bibinfo{author}{Rohani, P.},
  \bibinfo{author}{Bolker, B.~M.} \& \bibinfo{author}{Grenfell, B.~T.}
\newblock \bibinfo{journal}{\bibinfo{title}{{A simple model for complex
  dynamical transitions in epidemics}}}.
\newblock {\emph{\JournalTitle{Science}}} \textbf{\bibinfo{volume}{287}},
  \bibinfo{pages}{667--670} (\bibinfo{year}{2000}).

\bibitem{Stone2007}
\bibinfo{author}{Stone, L.}, \bibinfo{author}{Olinky, R.} \&
  \bibinfo{author}{Huppert, A.}
\newblock \bibinfo{journal}{\bibinfo{title}{{Seasonal dynamics of recurrent
  epidemics}}}.
\newblock {\emph{\JournalTitle{Nature}}} \textbf{\bibinfo{volume}{446}},
  \bibinfo{pages}{533--536} (\bibinfo{year}{2007}).
\newblock \doiprefix 10.1038/nature05638.

\bibitem{Diekmann1990}
\bibinfo{author}{Diekmann, O.}, \bibinfo{author}{Heesterbeek, J.} \&
  \bibinfo{author}{Metz, J.}
\newblock \bibinfo{journal}{\bibinfo{title}{{On the definition and the
  computation of the basic reproduction ratio R 0 in models for infectious
  diseases in heterogeneous populations}}}.
\newblock {\emph{\JournalTitle{Journal of Mathematical Biology}}}
  \textbf{\bibinfo{volume}{28}}, \bibinfo{pages}{365--382}
  (\bibinfo{year}{1990}).
\newblock \doiprefix 10.1007/BF00178324.

\bibitem{Hethcote1987}
\bibinfo{author}{Hethcote, H.~W.} \& \bibinfo{author}{{Van Ark}, J.~W.}
\newblock \bibinfo{journal}{\bibinfo{title}{{Epidemiological models for
  heterogeneous populations: proportionate mixing, parameter estimation, and
  immunization programs}}}.
\newblock {\emph{\JournalTitle{Mathematical Biosciences}}}
  \textbf{\bibinfo{volume}{84}}, \bibinfo{pages}{85--118}
  (\bibinfo{year}{1987}).
\newblock \doiprefix 10.1016/0025-5564(87)90044-7.

\bibitem{Nold1980}
\bibinfo{author}{Nold, A.}
\newblock \bibinfo{journal}{\bibinfo{title}{{Heterogeneity in
  disease-transmission modeling}}}.
\newblock {\emph{\JournalTitle{Mathematical Biosciences}}}
  \textbf{\bibinfo{volume}{52}}, \bibinfo{pages}{227--240}
  (\bibinfo{year}{1980}).
\newblock \doiprefix 10.1016/0025-5564(80)90069-3.

\bibitem{Hickson2014}
\bibinfo{author}{Hickson, R.~I.} \& \bibinfo{author}{Roberts, M.~G.}
\newblock \bibinfo{journal}{\bibinfo{title}{{How population heterogeneity in
  susceptibility and infectivity influences epidemic dynamics}}}.
\newblock {\emph{\JournalTitle{Journal of Theoretical Biology}}}
  \textbf{\bibinfo{volume}{350}}, \bibinfo{pages}{70--80}
  (\bibinfo{year}{2014}).
\newblock \doiprefix 10.1016/j.jtbi.2014.01.014.

\bibitem{Novozhilov2012}
\bibinfo{author}{Novozhilov, A.}
\newblock \bibinfo{journal}{\bibinfo{title}{{Epidemiological models with
  parametric heterogeneity: Deterministic theory for closed populations}}}.
\newblock {\emph{\JournalTitle{Mathematical Modelling of Natural Phenomena}}}
  \textbf{\bibinfo{volume}{7}}, \bibinfo{pages}{147--167}
  (\bibinfo{year}{2012}).
\newblock \doiprefix 10.1051/mmnp/20127310.

\bibitem{Keeling1999}
\bibinfo{author}{Keeling, M.~J.}
\newblock \bibinfo{journal}{\bibinfo{title}{{The effects of local spatial
  structure on epidemiological invasions}}}.
\newblock {\emph{\JournalTitle{Proceedings of the Royal Society B: Biological
  Sciences}}} \textbf{\bibinfo{volume}{266}}, \bibinfo{pages}{859--867}
  (\bibinfo{year}{1999}).
\newblock \doiprefix 10.1098/rspb.1999.0716.

\bibitem{Rohani1999}
\bibinfo{author}{Rohani, P.}
\newblock \bibinfo{journal}{\bibinfo{title}{{Opposite patterns of synchrony in
  sympatric disease metapopulations}}}.
\newblock {\emph{\JournalTitle{Science}}} \textbf{\bibinfo{volume}{286}},
  \bibinfo{pages}{968--971} (\bibinfo{year}{1999}).
\newblock \doiprefix 10.1126/science.286.5441.968.

\bibitem{Hagenaars2004}
\bibinfo{author}{Hagenaars, T.~J.}, \bibinfo{author}{Donnelly, C.~A.} \&
  \bibinfo{author}{Ferguson, N.~M.}
\newblock \bibinfo{journal}{\bibinfo{title}{{Spatial heterogeneity and the
  persistence of infectious diseases}}}.
\newblock {\emph{\JournalTitle{Journal of Theoretical Biology}}}
  \textbf{\bibinfo{volume}{229}}, \bibinfo{pages}{349--359}
  (\bibinfo{year}{2004}).
\newblock \doiprefix 10.1016/j.jtbi.2004.04.002.

\bibitem{Yu2009}
\bibinfo{author}{Yu, J.}, \bibinfo{author}{Jiang, D.} \& \bibinfo{author}{Shi,
  N.}
\newblock \bibinfo{journal}{\bibinfo{title}{{Global stability of two-group SIR
  model with random perturbation}}}.
\newblock {\emph{\JournalTitle{Journal of Mathematical Analysis and
  Applications}}} \textbf{\bibinfo{volume}{360}}, \bibinfo{pages}{235--244}
  (\bibinfo{year}{2009}).
\newblock \doiprefix 10.1016/j.jmaa.2009.06.050.

\bibitem{Colizza2006}
\bibinfo{author}{Colizza, V.}, \bibinfo{author}{Barrat, A.},
  \bibinfo{author}{Barth{\'{e}}lemy, M.} \& \bibinfo{author}{Vespignani, A.}
\newblock \bibinfo{journal}{\bibinfo{title}{{The role of the airline
  transportation network in the prediction and predictability of global
  epidemics}}}.
\newblock {\emph{\JournalTitle{Proceedings of the National Academy of
  Sciences}}} \textbf{\bibinfo{volume}{103}}, \bibinfo{pages}{2015--2020}
  (\bibinfo{year}{2006}).
\newblock \doiprefix 10.1073/pnas.0510525103.

\bibitem{Barthelemy2004}
\bibinfo{author}{Barth{\'{e}}lemy, M.}, \bibinfo{author}{Barrat, A.},
  \bibinfo{author}{Pastor-Satorras, R.} \& \bibinfo{author}{Vespignani, A.}
\newblock \bibinfo{journal}{\bibinfo{title}{{Velocity and hierarchical spread
  of epidemic outbreaks in scale-free networks}}}.
\newblock {\emph{\JournalTitle{Physical Review Letters}}}
  \textbf{\bibinfo{volume}{92}}, \bibinfo{pages}{178701--1}
  (\bibinfo{year}{2004}).
\newblock \doiprefix 10.1103/PhysRevLett.92.178701.

\bibitem{Keeling2005}
\bibinfo{author}{Keeling, M.~J.}
\newblock \bibinfo{journal}{\bibinfo{title}{{The implications of network
  structure for epidemic dynamics}}}.
\newblock {\emph{\JournalTitle{Theoretical Population Biology}}}
  \textbf{\bibinfo{volume}{67}}, \bibinfo{pages}{1--8} (\bibinfo{year}{2005}).
\newblock \doiprefix 10.1016/j.tpb.2004.08.002.

\bibitem{Hufnagel2004}
\bibinfo{author}{Hufnagel, L.}, \bibinfo{author}{Brockmann, D.} \&
  \bibinfo{author}{Geisel, T.}
\newblock \bibinfo{journal}{\bibinfo{title}{{Forecast and control of epidemics
  in a globalized world}}}.
\newblock {\emph{\JournalTitle{Proceedings of the National Academy of
  Sciences}}} \textbf{\bibinfo{volume}{101}}, \bibinfo{pages}{15124--15129}
  (\bibinfo{year}{2004}).
\newblock \doiprefix 10.1073/pnas.0308344101.

\bibitem{Holme2015}
\bibinfo{author}{Holme, P.}
\newblock \bibinfo{journal}{\bibinfo{title}{{Information content of
  contact-pattern representations and predictability of epidemic outbreaks}}}.
\newblock {\emph{\JournalTitle{Scientific Reports}}}
  \textbf{\bibinfo{volume}{5}}, \bibinfo{pages}{14462} (\bibinfo{year}{2015}).
\newblock \doiprefix 10.1038/srep14462.

\bibitem{Barthelemy2005}
\bibinfo{author}{Barth{\'{e}}lemy, M.}, \bibinfo{author}{Barrat, A.},
  \bibinfo{author}{Pastor-Satorras, R.} \& \bibinfo{author}{Vespignani, A.}
\newblock \bibinfo{journal}{\bibinfo{title}{{Dynamical patterns of epidemic
  outbreaks in complex heterogeneous networks}}}.
\newblock {\emph{\JournalTitle{Journal of Theoretical Biology}}}
  \textbf{\bibinfo{volume}{235}}, \bibinfo{pages}{275--288}
  (\bibinfo{year}{2005}).
\newblock \doiprefix 10.1016/j.jtbi.2005.01.011.

\bibitem{Boylan1991}
\bibinfo{author}{Boylan, R.~D.}
\newblock \bibinfo{journal}{\bibinfo{title}{{A note on epidemics in
  heterogeneous populations}}}.
\newblock {\emph{\JournalTitle{Mathematical Biosciences}}}
  \textbf{\bibinfo{volume}{105}}, \bibinfo{pages}{133--137}
  (\bibinfo{year}{1991}).
\newblock \doiprefix 10.1016/0025-5564(91)90052-K.

\bibitem{Andersson1998}
\bibinfo{author}{Andersson, H.} \& \bibinfo{author}{Britton, T.}
\newblock \bibinfo{journal}{\bibinfo{title}{{Heterogeneity in epidemic models
  and its effect on the spread of infection}}}.
\newblock {\emph{\JournalTitle{Journal of Applied Probability}}}
  \textbf{\bibinfo{volume}{35}}, \bibinfo{pages}{651--661}
  (\bibinfo{year}{1998}).
\newblock \doiprefix 10.1239/jap/1032265213.

\bibitem{Gardiner2003}
\bibinfo{author}{Gardiner, C.~W.}
\newblock \emph{\bibinfo{title}{{Handbook of stochastic methods}}}
  (\bibinfo{publisher}{Springer-Verlag}, \bibinfo{address}{Berlin Heidelberg},
  \bibinfo{year}{2003}), \bibinfo{edition}{3rd} edn.

\bibitem{vanKampen1992}
\bibinfo{author}{van Kampen, N.~G.}
\newblock \emph{\bibinfo{title}{{Stochastic processes in physics and
  chemistry}}} (\bibinfo{publisher}{Elsevier}, \bibinfo{address}{Amsterdam},
  \bibinfo{year}{1992}), \bibinfo{edition}{3rd} edn.

\bibitem{Rozhnova2009a}
\bibinfo{author}{Rozhnova, G.} \& \bibinfo{author}{Nunes, A.}
\newblock \bibinfo{journal}{\bibinfo{title}{{Cluster approximations for
  infection dynamics on random networks}}}.
\newblock {\emph{\JournalTitle{Physical Review E}}}
  \textbf{\bibinfo{volume}{80}}, \bibinfo{pages}{051915}
  (\bibinfo{year}{2009}).
\newblock \doiprefix 10.1103/PhysRevE.80.051915.

\bibitem{Kermack1927}
\bibinfo{author}{Kermack, W.~O.} \& \bibinfo{author}{McKendrick, A.~G.}
\newblock \bibinfo{journal}{\bibinfo{title}{{A contribution to the mathematical
  theory of epidemics}}}.
\newblock {\emph{\JournalTitle{Proceedings of the Royal Society A:
  Mathematical, Physical and Engineering Sciences}}}
  \textbf{\bibinfo{volume}{115}}, \bibinfo{pages}{700--721}
  (\bibinfo{year}{1927}).
\newblock \doiprefix 10.1098/rspa.1927.0118.

\bibitem{Britton2002}
\bibinfo{author}{Britton, T.} \& \bibinfo{author}{O'Neill, P.~D.}
\newblock \bibinfo{journal}{\bibinfo{title}{{Bayesian Inference for Stochastic
  Epidemics in Populations with Random Social Structure}}}.
\newblock {\emph{\JournalTitle{Scandinavian Journal of Statistics}}}
  \textbf{\bibinfo{volume}{29}}, \bibinfo{pages}{375--390}
  (\bibinfo{year}{2002}).
\newblock \doiprefix 10.1111/1467-9469.00296.

\bibitem{Shulgin1998}
\bibinfo{author}{Shulgin, B.}, \bibinfo{author}{Stone, L.} \&
  \bibinfo{author}{Agur, Z.}
\newblock \bibinfo{journal}{\bibinfo{title}{{Pulse vaccination strategy in the
  SIR epidemic model.}}}
\newblock {\emph{\JournalTitle{Bulletin of mathematical biology}}}
  \textbf{\bibinfo{volume}{60}}, \bibinfo{pages}{1123--1148}
  (\bibinfo{year}{1998}).
\newblock \doiprefix 10.1016/S0092-8240(98)90005-2.

\bibitem{Gillesple1977}
\bibinfo{author}{Gillespie, D.~T.}
\newblock \bibinfo{journal}{\bibinfo{title}{{Exact stochastic simulation of
  coupled chemical reactions}}}.
\newblock {\emph{\JournalTitle{The Journal of Physical Chemistry}}}
  \textbf{\bibinfo{volume}{81}}, \bibinfo{pages}{2340--2361}
  (\bibinfo{year}{1977}).
\newblock \doiprefix 10.1021/j100540a008.

\bibitem{Keeling2008}
\bibinfo{author}{Keeling, M.~J.} \& \bibinfo{author}{Rohani, P.}
\newblock \emph{\bibinfo{title}{{Modeling Infectious Diseases in Humans and
  Animals}}} (\bibinfo{publisher}{Princeton University Press},
  \bibinfo{address}{Princeton, NJ}, \bibinfo{year}{2008}).

\bibitem{Anderson1992}
\bibinfo{author}{Anderson, R.~M.} \& \bibinfo{author}{May, R.~M.}
\newblock \emph{\bibinfo{title}{{Infectious diseases of humans: Dynamics and
  control}}} (\bibinfo{publisher}{Oxford University Press},
  \bibinfo{address}{Oxford}, \bibinfo{year}{1992}).

\bibitem{Stoica2004a}
\bibinfo{author}{Stoica, P.} \& \bibinfo{author}{Moses, R.}
\newblock \emph{\bibinfo{title}{{Spectral analysis of signals}}}
  (\bibinfo{publisher}{Pearson Prentice Hall}, \bibinfo{address}{Upper Saddle
  River}, \bibinfo{year}{2004}).

\bibitem{Ball1985}
\bibinfo{author}{Ball, F.}
\newblock \bibinfo{journal}{\bibinfo{title}{{Deterministic and stochastic
  epidemics with several kinds of susceptibles}}}.
\newblock {\emph{\JournalTitle{Advances in Applied Probability}}}
  \textbf{\bibinfo{volume}{17}}, \bibinfo{pages}{1} (\bibinfo{year}{1985}).
\newblock \doiprefix 10.2307/1427049.

\bibitem{Yates2006}
\bibinfo{author}{Yates, A.}, \bibinfo{author}{Antia, R.} \&
  \bibinfo{author}{Regoes, R.~R.}
\newblock \bibinfo{journal}{\bibinfo{title}{{How do pathogen evolution and host
  heterogeneity interact in disease emergence?}}}
\newblock {\emph{\JournalTitle{Proceedings of the Royal Society B: Biological
  Sciences}}} \textbf{\bibinfo{volume}{273}}, \bibinfo{pages}{3075--3083}
  (\bibinfo{year}{2006}).
\newblock \doiprefix 10.1098/rspb.2006.3681.

\bibitem{Butler2009}
\bibinfo{author}{Butler, T.} \& \bibinfo{author}{Goldenfeld, N.}
\newblock \bibinfo{journal}{\bibinfo{title}{{Robust ecological pattern
  formation induced by demographic noise}}}.
\newblock {\emph{\JournalTitle{Physical Review E}}}
  \textbf{\bibinfo{volume}{80}}, \bibinfo{pages}{030902}
  (\bibinfo{year}{2009}).
\newblock \doiprefix 10.1103/PhysRevE.80.030902.

\bibitem{Black2012}
\bibinfo{author}{Black, A.~J.} \& \bibinfo{author}{McKane, A.~J.}
\newblock \bibinfo{journal}{\bibinfo{title}{{Stochastic formulation of
  ecological models and their applications}}}.
\newblock {\emph{\JournalTitle{Trends in Ecology {\&} Evolution}}}
  \textbf{\bibinfo{volume}{27}}, \bibinfo{pages}{337--345}
  (\bibinfo{year}{2012}).
\newblock \doiprefix 10.1016/j.tree.2012.01.014.

\bibitem{Cremer2008}
\bibinfo{author}{Cremer, J.}, \bibinfo{author}{Reichenbach, T.} \&
  \bibinfo{author}{Frey, E.}
\newblock \bibinfo{journal}{\bibinfo{title}{{Anomalous finite-size effects in
  the Battle of the Sexes}}}.
\newblock {\emph{\JournalTitle{The European Physical Journal B}}}
  \textbf{\bibinfo{volume}{63}}, \bibinfo{pages}{373--380}
  (\bibinfo{year}{2008}).
\newblock \doiprefix 10.1140/epjb/e2008-00036-x.

\bibitem{Mobilia2010}
\bibinfo{author}{Mobilia, M.}
\newblock \bibinfo{journal}{\bibinfo{title}{{Oscillatory dynamics in
  rock–paper–scissors games with mutations}}}.
\newblock {\emph{\JournalTitle{Journal of Theoretical Biology}}}
  \textbf{\bibinfo{volume}{264}}, \bibinfo{pages}{1--10}
  (\bibinfo{year}{2010}).
\newblock \doiprefix 10.1016/j.jtbi.2010.01.008.

\bibitem{Rozhnova2012}
\bibinfo{author}{Rozhnova, G.}, \bibinfo{author}{Nunes, A.} \&
  \bibinfo{author}{McKane, A.~J.}
\newblock \bibinfo{journal}{\bibinfo{title}{{Phase lag in epidemics on a
  network of cities}}}.
\newblock {\emph{\JournalTitle{Physical Review E}}}
  \textbf{\bibinfo{volume}{85}} (\bibinfo{year}{2012}).
\newblock \doiprefix 10.1103/PhysRevE.85.051912.

\end{thebibliography}
\end{document}